\documentclass[preprint]{aastex} 
\usepackage{epsf,natbib,latexsym}
\newcommand{\skatrue}{ S_{31, T}}
\newcommand{\skaobs}{ S_{31, obs}}
\newcommand{\sltrue}{ S_{1.4, T}}
\newcommand{\slobs}{ S_{1.4, obs}}
\newcommand{\sratio}{\frac{S_{31,T}}{S_{1.4,T}}}
\newcommand{\jysq}{\, {\rm Jy^2/Sr}}
\newcommand{\mJy}{\,{\rm mJy} }
\newcommand{\K}{\,{\rm K} }
\newcommand{\uJy}{\,{\rm \mu Jy} }
\newcommand{\uKbp}{\,{\rm \mu K^2} }

\newcommand{\Hz}{\, {\rm Hz} }

\newcommand{\ie}{{\it i.e.}}

\newcommand{\ghz}{{\, \rm GHz}}

\newcommand{\nK}{\,{\rm n K} }

\begin{document}

\title{A 31 GHz Survey of Low-Frequency Selected Radio Sources}

\author{
 B.S. Mason\altaffilmark{1},
 L. Weintraub\altaffilmark{2}, 
 J. Sievers\altaffilmark{3}, 
 J.R. Bond\altaffilmark{3},
 S.T. Myers\altaffilmark{4},
 T.J. Pearson\altaffilmark{2}, 
 A.C.S. Readhead\altaffilmark{2},
 M. C. Shepherd\altaffilmark{2}
}

\altaffiltext{1}{National Radio Astronomy Observatory, 520 Edgemont Road, Charlottesville, VA 22903}
\altaffiltext{2}{Owens Valley Radio Observatory, California Institute of
Technology, Pasadena, CA}
\altaffiltext{3}{Canadian Institute for Theoretical
Astrophysics, University of Toronto, ON M5S 3H8, Canada} 
\altaffiltext{4}{National Radio Astronomy Observatory, Soccorro, NM 87801} 

\begin{abstract}

The 100-m Robert C. Byrd Green Bank Telescope (GBT) and the 40-m Owens
Valley Radio Observatory (OVRO) telescope have been used to conduct a
31 GHz survey of $3,165$ known extragalactic radio sources over $ 143
\, {\rm deg^2}$ of the sky. Target sources were selected from the NRAO
VLA Sky Survey in fields observed by the Cosmic Background Imager
(CBI); most are extragalactic active galactic nuclei (AGN) with $1.4$
GHz flux densities of 3 to 10 mJy. The resulting 31 GHz catalogs are
presented in full online.  Using a Maximum-Likelihood analysis to
obtain an unbiased estimate of the distribution of the $1.4$ to $31$
GHz spectral indices of these sources, we find a mean $31$ to $1.4$
GHz flux ratio of $0.110 \pm 0.003$ corresponding to a spectral index
of $\alpha=-0.71 \pm 0.01 $ ($S_{\nu} \propto \nu^{\alpha}$); $ 9.0
\pm 0.8 \%$ of sources have $\alpha > -0.5$ and $ 1.2 \pm 0.2 \%$ have
$\alpha > 0$.  By combining this spectral index distribution with
$1.4$ GHz source counts we predict 31 GHz source counts in the range
$1 \mJy < S_{31} < 4 \mJy$, $N(>S_{31}) = (16.7 \pm 1.7) \, {\rm
deg^{-2}} (S_{31}/{\rm 1 \, mJy})^{-0.80 \pm 0.07}$.  We also assess
the contribution of mJy-level ($S_{\rm 1.4 \, GHz} < 3.4 \mJy$) radio
sources to the 31 GHz CMB power spectrum, finding a mean power of
$\ell (\ell+1) C^{src}_{\ell}/(2\pi) = 44 \pm 14 \uKbp$ and a $95\%$
upper limit of $80 \uKbp$ at $\ell = 2500$.  Including an estimated
contribution of $12 \uKbp$ from the population of sources responsible
for the turn-up in counts below $S_{\rm 1.4 \, GHz} = 1 \mJy$ this
amounts to $21 \pm 7 \%$ of what is needed to explain the CBI
high-$\ell$ excess signal, $275 \pm 63 \uKbp $. These results are
consistent with other measurements of the 31 GHz point source
foreground.

\end{abstract}


\section{Introduction}
\label{sec:intro}

Accurate measurements of cosmic microwave background (CMB)
anisotropies on scales smaller than the sound horizon at last
scattering are necessary to form a complete understanding of the
initial power spectrum of cosmic inhomogeneities.  Secondary
anisotropies--- chiefly reionization and Sunyaev-Zel'dovich (SZ)
distortions from large scale structures--- also leave imprints on
these scales which contain clues of the subsequent evolution of
structures from very simple, linear initial conditions to the wealth
of nonlinear structures seen in the nearby universe.  Photon diffusive
damping occuring near last scattering has the effect of strongly
suppressing small-scale intrinsic CMB anisotropies, so precise
measurements on these scales are difficult. A key challenge for
centimeter-wavelength observations on these angular scales is the
foreground presented by faint, discrete radio sources.

The Cosmic Background Imager \citep[CBI]{cbi7} is a 30 GHz, small
angular-scale experiment which has detected, at $2.9 \sigma$, power in
excess of intrinsic CMB anisotropy at multipoles $\ell > 2000$. The
accuracy of this result is limited by our knowledge of the
contribution of faint extragalactic radio sources to the measured
power spectrum. The CBI removes all {\it known} radio sources by
``projecting'' them out of the dataset\footnote{``Source projection''
is a procedure equivalent to allowing an arbitrary variance to a
linear combination of the data corresponding to the point source. This
procedure is performed in fourier space; the analog in the image
domain would be deleting a contaminated pixel.}. The best available
low-frequency radio data covering the CBI fields are from the $1.4 \,
{\rm GHz}$ NRAO VLA Sky Survey \citep[NVSS][]{nvss}, which is reliable
and complete down to $3.4 \mJy$. Since the source density on the sky
as a function of flux density (the ``source counts'') is well known to
down to $50-100 \uJy$, the 31 GHz power spectrum of sources not
detected in the NVSS can be calculated given sufficient knowledge of
the ratio of flux densities at these two frequencies,
$S_{30}/S_{1.4}$. Since the sources are unresolved they will have a
flat power spectrum: $C_{\ell} \sim {\rm const}$, or $\ell (\ell +1)
C_{\ell} \sim \ell^2$ in terms of the prevailing convention for CMB
power spectra. The power spectrum of these sources is directly
subtracted from the CBI power spectrum. In this paper we call this
correction to the power spectrum the ``residual source correction'';
in other papers it is also called the statistical correction and the
isotropic point source correction.  The uncertainty in this correction
is comparable to other uncertainties on the smallest angular scales
measured by CBI ($\ell > 2000$, or angular scales smaller than about 5
arcminutes).  \citet{cbi7} calculate a value of $96 \pm 48 \uKbp$,
where the uncertainty in the correction is dominated by the poorly
constrained extrapolation of $1.4 \, {\rm GHz}$ flux densities to $31
\, {\rm GHz}$. The $\sim 50 \uKbp$ uncertainty in the point source
correction contributes a substantial uncertainty to the CBI
measurement at $\ell > 1960$ of $355^{+137}_{-122} \uKbp$ (of which
$80 - 90 \uKbp$ is expected to be intrinsic anisotropy). New CBI
results presented in \citet{cbi10} show an excess which, to be
explained by point sources, would have a power $\ell
(\ell+1)C^{src}_{\ell}/(2\pi) = 275 \pm 63 \uKbp$ at $\ell =2500$,
consistent with the \citet{cbi7} result.

In more detail, if at a given frequency sources above some flux
density $S_{max}$ are dealt with, the power spectrum of
residual sources below $S_{max}$
is proportional to
\citep{myers03}
\begin{equation}
X_{src} = \int_{0}^{S_{max}} \, dS \, S^2 \frac{dN}{dS}
\label{eq:csrc}
\end{equation}
In our case the sources are selected for inclusion at a
different, lower frequency, ({\it i.e.}, they are in effect the
sources in our fields that are not detected reliably in the NVSS,
hence are not projected out).  Then, assuming that the spectral
properties of the sources are constant over the range of relevant flux
densities--- in practice, $\sim 1$ to $4$ mJy at $1.4$ GHz---
Eq.~\ref{eq:csrc} becomes
\begin{equation}
X_{src} = \langle \left(\frac{S_{31}}{S_{1.4}}\right)^2 \rangle \int_{0}^{S_{max,1.4}} \, dS_{1.4} \, 
S_{1.4}^2 \, \frac{dN}{dS_{1.4}}
\end{equation}
The low-frequency source counts $dN/dS$ are well-known.  To accurately
correct 31 GHz CMB measurements for point source contamination we need
to determine the mean value of $ (S_{31}/S_{1.4})^2 $, and to
determine the uncertainty in $X_{src}$ and therefore in the
source-corrected CMB power spectrum, we must know its distribution.
For the shallow source counts [$N(>S) \propto S^{-0.7}$] observed at
mJy levels the sources immediately below the projection threshold
dominate the sky variance and our surveys target this population. For
the CBI, for instance, 75\% of the correction comes from sources in
the range $1 < S_{1.4} < 3.4 \mJy$.  At 31 GHz the sky variance can be
strongly influenced by the abundance of comparatively rare flat or
inverted spectrum sources, so a large sample is needed to place useful
constraints on the abundance of these sources.

In this paper we present a detailed characterization of the impact of
the discrete source foreground on arcminute-scale 31 GHz anisotropy
measurements based upon two observational campaigns. The first
campaign was carried out with the Owens Valley Radio Observatory
(OVRO) 40-meter telescope at 31 GHz from September 2000 through
December 2002.  The second campaign used the Robert C. Byrd
Green Bank Telescope (GBT) from February to May of 2006.  This work
was undertaken with the specific aim of improving the accuracy of CBI
microwave background anisotropy measurements. A companion paper
\citep{cbi10} presents the 5-year CBI total intensity power spectrum
incorporating the results of the point source measurements discussed
here. 

The structure of this paper is as follows. In \S~\ref{sec:instr} we
describe the instrumentation used in both surveys. \S~\ref{sec:samp}
describes the source lists and sample selection, and
\S~\ref{sec:obsreduc} describes the OVRO 40-m and GBT observations and
data reduction and presents catalogs of the
observations. \S~\ref{sec:results} presents a determination of the
$1.4$ to $31$ GHz spectral index of NVSS sources and determines the
implications of these measurements for 31 GHz CMB observations, for
the case of the CBI in particular; here we also present a
determination of the 31 GHz source counts.  Finally
\S~\ref{sec:summary} reviews our main conclusions.


\section{Instrumentation}
\label{sec:instr}

\subsection{OVRO 40-meter and 31 GHz Receiver}
\label{subsec:ovroinstr}

In 1999 the OVRO 40-meter telescope was outfitted with a
Dicke-switching, dual horn receiver operating in four $2 \ghz$ bands
between 26 and 34 GHz.  The Dicke switch alternates between the two
horns at a rate of $125 \Hz$, sampling two beams separated by $7'.8$
in cross-elevation on the sky.  Each beam has a FWHM of $1'.36$ at 31
GHz; this is somewhat larger than what would be expected for a
40-meter dish since only the central 30 meters are illuminated.  The
measured receiver temperature in the 31 GHz band is $23 \K$. The
statistics of (noise) samples taken against both ambient temperature
and $77 \K$ beam-filling loads are consistent with the receiver
temperature.  Including $ 13 \K$ per airmass due to the atmosphere,
$2.7 \K$ from the CMB, and a fixed ground contribution of $ 10 \K$,
the system temperature at zenith is $\sim 50 \K$.  Calibration is
facilitated by two broad-band noise diodes; cross-guide couplers
before the Dicke switch allow for the insertion of signals from these
devices.

The telescope is an alt-azimuth instrument consisting of a
paraboloidal dish reflector, primary focus feed, and supports, mounted
on an alidade and base pedestal.  Our observing frequency of $ 31
\ghz$ is beyond the design specification of the 40-meter telescope,
resulting in aperture efficiencies of only $15\%$.  The gain of the
40-m changes as a function of both zenith-angle (ZA) due to
gravitational deformations in the dish structure, and of angle of the
sun from the optical axis due to thermal deformations.  These
variations were characterized by long tracks on bright calibrators and
the resulting gain corrections are applied offline in the data
reduction.  The focus position also varies with zenith angle.  The
focus position corrections are applied in real time and checked
periodically for consistency.  The peak sensitivity that the 40-meter
achieves is $12$ mJy RMS in one minute at 40 degrees elevation in the
31 GHz band.  The outer bands had substantially higher noise
levels. The analysis in this paper relies upon data from the 31 GHz
channel only.


\subsection{GBT, 31 GHz Receiver and Continuum Backend}
\label{subsec:gbtinstr}

The Robert C. Byrd Green Bank Telescope (GBT) is a 100-meter off-axis
Gregorian telescope located in Green Bank, West
Virginia \citep{gbtspie}. One of the GBT's distinguishing features is
the fact that it was designed to operate efficiently at frequencies up
through 115 GHz. For the observations presented here the GBT's 31 GHz
aperture efficiency was 50\%.  The 2-dimensional RMS referenced
pointing accuracy is $4''$ on timescales of half an hour to an
hour. Corrections to the focus tracking model on the same timescale
are typically a few millimeters.  Owing to the primary reflector's
remotely actuated positioning system, the telescope gain does not vary
significantly at elevations greater than 20 degrees.

Broadband measurements of the radio-frequency continuum are affected
by systematic effects such as gain fluctuations and variation in the
emissivity of the atmosphere. To this end the receiver built for the
OVRO 40-m employed a Dicke switch, enabling two feeds to be sampled
rapidly in series. For the GBT receiver, we chose an electronic
beamswitching arrangement employing $180^{\circ}$ hybrids similar to
the WMAP radiometers \citep{jarosik, padinka}.  This permits  wider
bandwidth coverage, avoids expensive and difficult to procure
Dicke switches, and permits signals from both feeds to be
simultaneously measured at all times. The receiver provides 16
continuum channels, one for each of four frequency bands, two feeds,
and both circular polarizations.  Receiver temperatures range from $20
\, {\rm K}$ to $40 \, {\rm K}$ across the band, resulting in $T_{sys}$
values on-sky of $35 \, {\rm K}$ to $65 \, {\rm K}$.  Broadband noise
diodes are coupled in to the signal path prior to the first hybrid tee
and permit monitoring the total gain of the system for calibration
purposes.

The Caltech Continuum Backend (CCB) is a digital backend which
controls the GBT 26-40 GHz receiver and synchronously reads out and
demodulates the beamswitched signal.  Lab tests conducted with the CCB
connected to the receiver show that the noise obtained is 15 to 30\%
above the theoretical noise minimum given by the radiometer equation.
For our observations we operated the receiver with a 4 kHz beam switch
rate in order to avoid excessive lost of time to blanking after phase
switch transitions. The CCB was constructed by NRAO and the California
Institute of Technology, with funding provided to Caltech through the
NRAO University Instrumentation Program. The 26-40 GHz receiver and
CCB are available as facility instruments on the GBT.


With the 31 GHz receiver and continuum backend, we attain nearly
thermal noise limited performance ($\sim 0.15 \mJy$ RMS in one minute)
a small fraction of the time, when the atmosphere is very dry and
stable. More typically the thermal noise RMS in a 60 second
double-differenced observation (essentially identical to that
described in \S~\ref{sec:ovroobs}) is $0.4 - 0.5 \mJy$ (RMS). The
results in this paper depend only on data from the 31 GHz channels.

\section{Source Lists and Sample Selection}
\label{sec:samp}

The CBI total intensity mosaic fields \citep{cbi7} covered $98 \, {\rm
deg^2}$ of sky, including a $45'$ buffer zone. The OVRO 40m survey
targeted all $S_{1.4} > 6 {\rm mJy}$ sources in this region. The CBI
polarization observations \citep{cbipol} covered $115 \, {\rm deg^2}$
in all, also including a $45'$ buffer zone; accounting for the overlap
between these two datasets the total sky coverage is $143 \, {\rm
  deg^2}$. The GBT survey targeted $S_{1.4} > 3.4 \mJy$ (the NVSS 99\%
completeness limit) sources in this total region, although full
coverage was not achieved.  Source selection proceeded from areas of
sky with the lowest CBI map noise. Sources detected at $3\sigma$ or
greater in the OVRO survey as a rule were avoided in the GBT survey.





Sources in the CBI fields were observed from September 1999 through
December 2001 with the OVRO 40-meter telescope in support of ongoing
CBI observations. The 40-m observations preferentially targetted
sources in the original \citep{cbi3,cbi7} CBI total intensity fields;
in all, $2,315$ sources were observed by the 40-m. With the typical
RMS sensitivity of the OVRO survey ($2.5 \, {\rm mJy}$ -- see
\S~\ref{sec:ovroobs}), this resulted in 180 detections at $4\sigma$ or
greater significance (363 at $3\sigma$ or greater).

\begin{figure}[htbp]
\vspace{3in}
\plotone{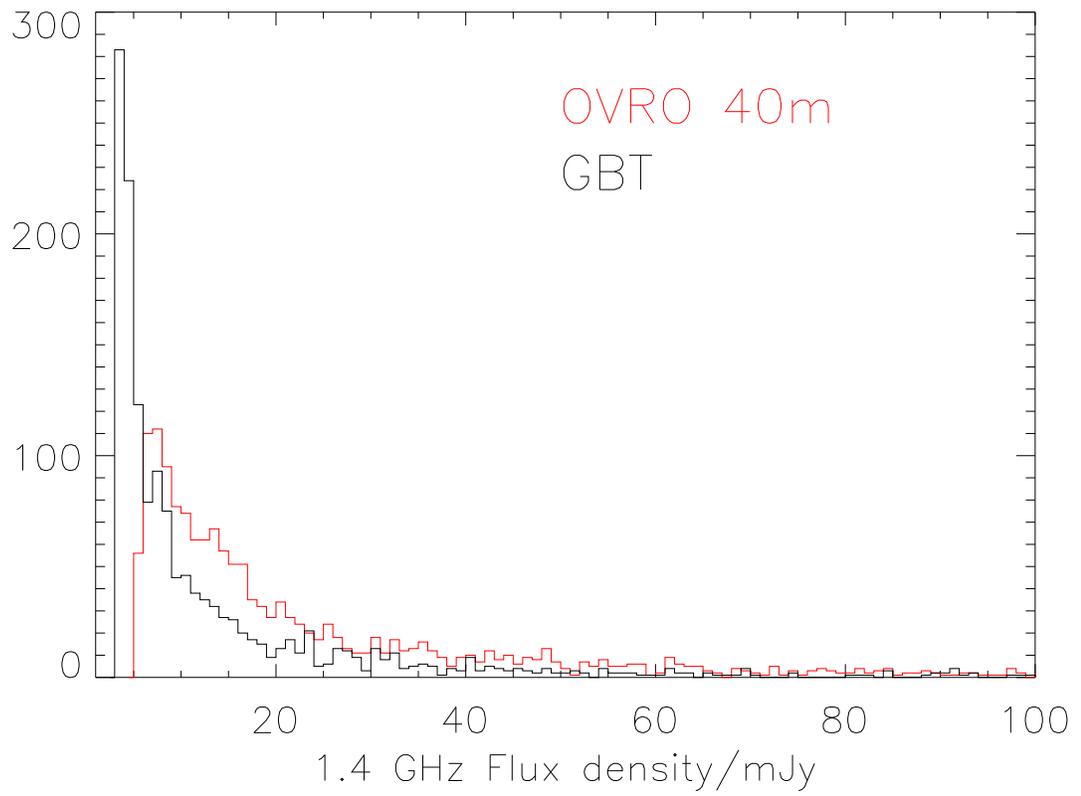}
\caption{NVSS flux densities for sources measured in the GBT and OVRO 40-m surveys.}
\label{fig:sourcefluxdist}
\end{figure}

Our aim with the GBT survey was to measure {\it all} of the NVSS
sources in the CBI fields with a sensitivity comparable to the RMS
noise in the CBI maps.  The 363 sources previously
detected\footnote{Due to a software bug, GBT observations of sources
  between $+01^{\circ}$ and $-01^{\circ}$ were observed without regard
  for the OVRO 40m measurements, {\it i.e.}  observations in this
  range were not pre-censored by the measured OVRO flux value.} at
$3\sigma$ or greater in the OVRO 40m survey were not observed by the
GBT, leaving $5,636$ sources. Useful GBT data were collected on
$1,490$ of these (\S~\ref{sec:gbtobs}). Faint NVSS sources, and
sources in areas where the CBI maps are most sensitive, were
preferntially targeted.  Of the OVRO observations, 640 sources' data
were superceded by more sensitive GBT measurements, leaving $1675$
unique observations in the OVRO dataset. In all useful data were
obtained on $3,165$ NVSS sources.  The distribution of $1.4$ GHz flux
densities of the sources measured is shown in
Figure~\ref{fig:sourcefluxdist}.



\section{Observations}
\label{sec:obsreduc}

\subsection{OVRO Observations}
\label{sec:ovroobs}

All NVSS sources brighter than 6 mJy were observed to a typical RMS
sensitivity of $2.4 \mJy$, requiring $ 30$ five minute observations on
average.  Five minute observations of NVSS-selected sources were
interleaved with daily measurements of 3C286, 3C147, 3C279, and other
bright sources  to monitor the system's
performance.  In between calibrator observations, the system gain was
monitored with the noise diodes internal to the receiver.  Every 40
minutes, a calibrator source within $\sim 15^{\circ}$ of the field
being observed was measured to determine the telescope pointing
offset.  For each flux density measurement, the online system reports
an estimate of the measurement error from the variance in the 1-second
samples within the integration period.  During the course of all
observations weather data are collected and recorded for later
correlation with the astronomical datastream.  The basic measurement
consists of four beamswitched integration periods wherein the
source of interest is alternately placed in the two beams of the
telescope in an A/B/B/A pattern. This symmetric double-differencing
scheme is effective at cancelling constant and gradient terms in
atmospheric emission \citep{readheadovro}.

Our flux density scale is based on the WMAP 5-year 
measurement of the 32 GHz brightness temperature of Jupiter $T_J =
146.6 \pm 0.75 \K$ \citep{hill}, extrapolated across the receiver bandpass with
3C286 using an assumed spectral index of $\alpha = -0.827$
\citep{ring5m}.  Observations of 3C286, 3C48, and 3C147 from Sep 1999
through May of 2000 showed uncorrelated RMS variations in flux density
of $\sim 4 \%$.  Together with the $3\%$ uncertainty in the absolute
flux density of Jupiter, this gives a $5\%$ calibration uncertainty
for the OVRO 40-m.

\subsection{OVRO Data Reduction}
\label{sec:ovroreduc}

OVRO data were reduced as follows.  Observations of the noise diodes
several times per hour are first used to remove electronic gain
fluctuations and scale the data to antenna temperature. Corrections
for the telescope gain as a function of elevation, determined from
long tracks on bright calibrators, are applied, as is a correction to
account for atmospheric opacity as a function of elevation.  


Time-variable weather conditions determine the sensitivity of OVRO
40-m flux density measurements. Typically sources were observed in
rotation, with 5 1-minute observations of each source. The 
timescale on which changes in the weather affect the 40-m photometric
sensitivity is typically an hour, which sets the timescale on which we wish to
estimate the measurement noise. During this period of time $\sim 10$
separate sources with different mean flux densities were measured.
Sources in the OVRO sample are expected to have 31 GHz flux
densities of $1-2$ mJy, small in comparison to the $10-15$ mJy noise
level achieved on a per-observation basis.  To determine the
measurement noise for a single observation, we wished to compute the
characteristic scatter of similar observations nearby in time with no
contribution due to the scatter in the source flux densities
themselves, our main concern being the power-law tail of the source
flux density distribution which results in rare objects comparable to
or greater than the per-observation noise level which could bias a
scatter-based noise estimate.  We chose to determine the noise level
of each observation by computing the median absolute deviation of
similar observations within a one hour buffer centered on the
observation under consideration. The median absolute deviation (MedAD)
of some data $x_i$ is defined as
\begin{equation}
 {\rm MedAD} (x_i) = {\rm Median} ( | x_i - {\rm Median}(x_i) | )
\end{equation}
and is a measure of the dominant noise scale of a distribution which
is extremely resistant to the presence and magnitude of outliers
\citep{robust}.  For a Gaussian distribution of width $\sigma_g$, the
${\rm MedAD} = 0.6745 \sigma_g$. We used this relation to to rescale
the MedAD into an effective Gaussian $\sigma$ value for each
observation.  An indepenent noise estimate was provided by the scatter
of 1-second integrations {\it within} each measurement. The internal
noises gave results comparable to, but generally 15\%-30\% lower than,
the scatter between observations. This is consistent with the
expectation that except in the very best observing conditions,
low-level photometric instabilities ({\it e.g.}  a slow variation in
the gradient of the sky emission) on timescales longer than that of an
individual source observation contribute to the measurement noise.  We
confirmed with simulations that to within a few percent this approach
yields an unbiased estimate of the RMS of a Gaussian noise
distribution in the presence of power-law source populations with
typical per-measurement signal-to-noise ratios of $0.1$ to $0.2$.

\begin{figure}
\vspace{4in}
\includegraphics{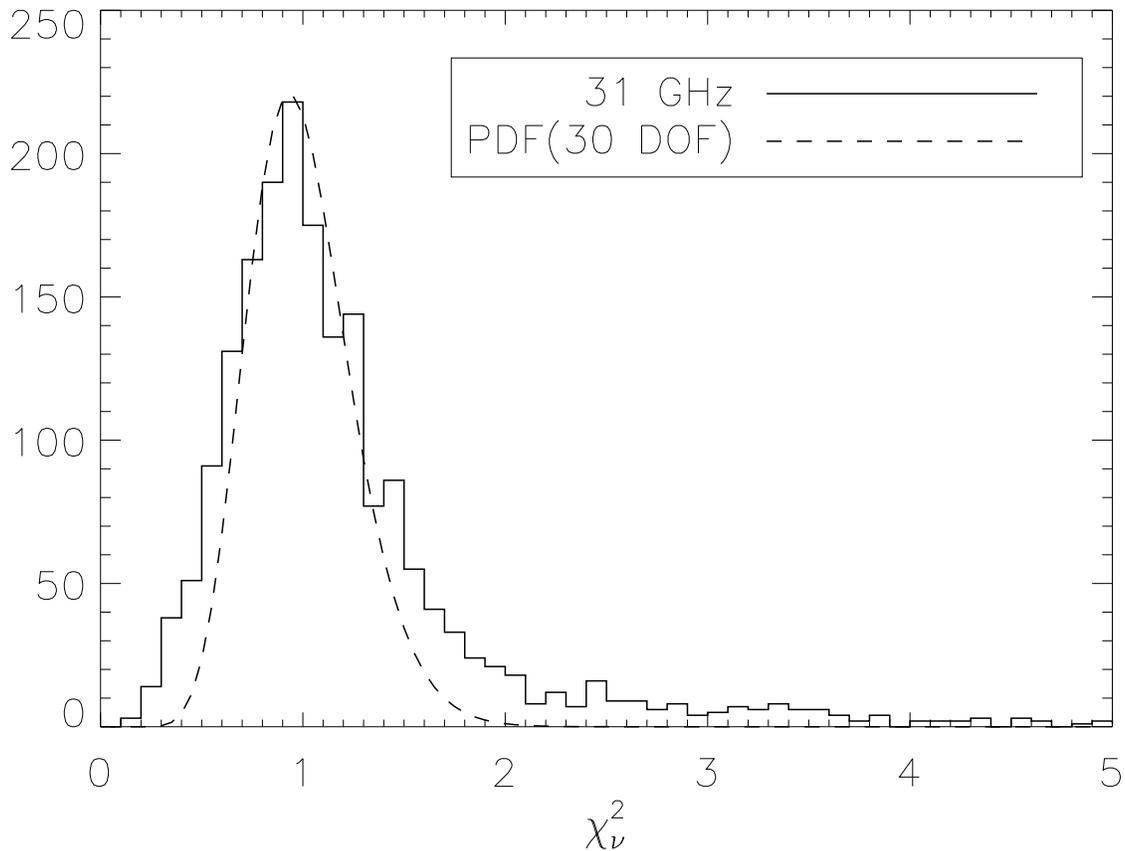}
\caption{The $\chi^2_{\nu}$ distribution for OVRO 40m measurements, each individual
$\chi^2_{\nu}$ value being derived from the combination of all data in the given
frequency band on a given source. Also shown is the theoretical reduced $\chi^2$
distribution for 30 degrees of freedom, which is typical, although the number of
observations on a given source ranges from three to a hundred in some cases.}
\label{fig:ovrochi2}
\end{figure}

\begin{figure}
\vspace{4in}
\includegraphics{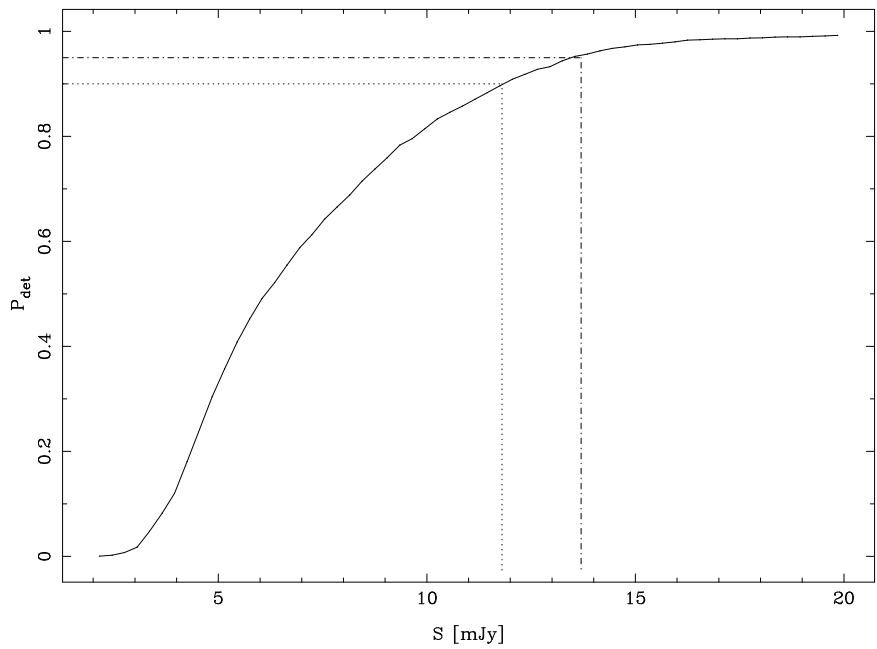}
\caption{The probability that a given source is detected in the OVRO
survey as a function of 31 GHz flux density.  The 90\% (dotted) and
95\% (dash-dotted) completeness levels are shown for clarity. The GBT survey
is essentially complete above $2.5 \mJy$.}
\label{fig:ovrosens}
\end{figure}

Over the course of the observing campaign individual sources in our
sample were observed between a few and about 100 times, most commonly
about thirty times.  All observations of a given source were combined
to form a weighted mean and an uncertainty on this mean was computed
by propagation of the error bars of each measurement. The reduced
$\chi^2$ of the data about this average was also computed; these
values are shown in Figure~\ref{fig:ovrochi2}.  For reference the
expected distribution of $\chi^2_{\nu}$ for 30 DOF is shown.  There is
a fairly broad distribution of final sensitivies, owing largely to the
range of total integration times per source, but also to the range of
photometric conditions. For our final source catalog
(\S~\ref{subsec:catalogs}) we adopted a $4\sigma$ detection threshold;
the probability of detecting a random source of a given flux density,
allowing for the distribution of noises in our dataset, is shown in
Figure~\ref{fig:ovrosens}.

We reject data within 5 degrees of the sun or the moon, as well as
data for which the preceding pointing calibrator observations were not
successful or did not occur within one hour. Since high winds can
affect the telescope pointing, observations during which the wind was
greater than 15 mph are discarded.  Reducing the pointing requirement
from an hour to a half-hour, and the wind limit from 15 to $7.5$ mph
(reducing the force of the wind on the telescope by roughly a factor
of four) did not significantly affect the measured flux densities of
the OVRO-detected sources.

The mean flux density for each source is computed by averaging over
the entire observing epoch, with each point inversely weighted by the
estimated measurement variance.  For our final results we used only the
$30-32$ GHz band, which we take to have a nominal center frequency of
$31.0$ GHz. This is the center of the CBI observing band and
corresponds closely to the most sensitive channel of the GBT receiver,
readily allowing direct comparisons.

\subsection{GBT Observations}
\label{sec:gbtobs}

Test observations of the CCB and 26-40 GHz receiver on the GBT were
conducted in November and December of 2005, and January of 2006; these
observations confirmed lab measurements of the system performance.
The science observations ran from 02 February, 2006 through 07 May
2006.  We collected 3198 observations of 3040 NVSS sources; after the
data filtering described below, 1567 observations of 1490 sources
remain in the final dataset.

\begin{figure}[htbp]
\epsfxsize=6.2in
\epsffile{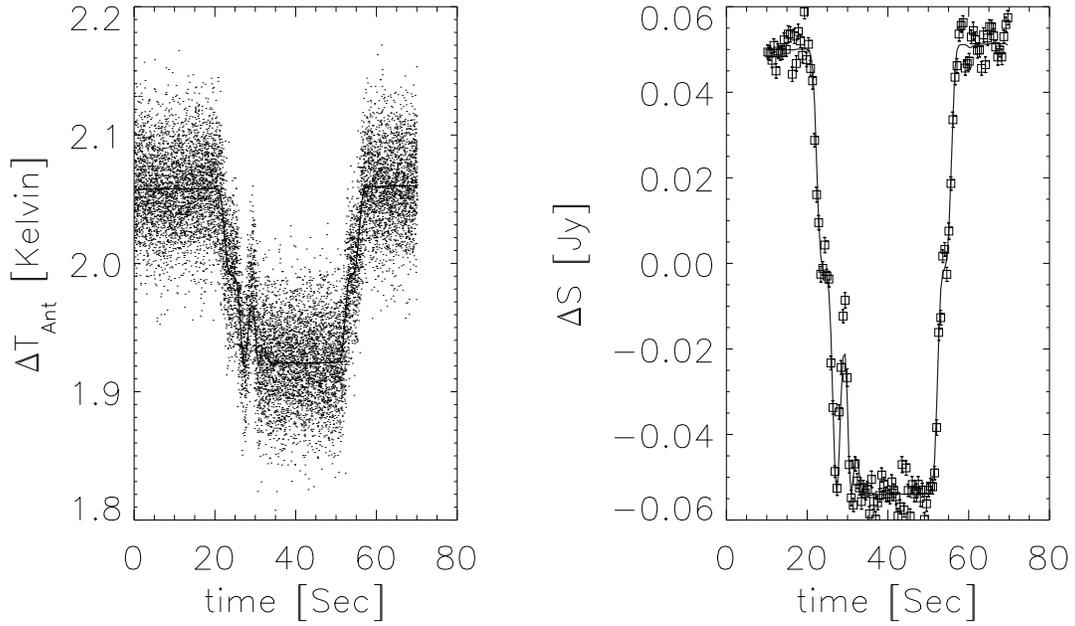}
\caption{Measurement of steep spectrum source $085328-0341$ at $31.25
  \, {\rm GHz}$.  The left hand plot shows the individual 5
  millisecond beamswitched integration calibrated via the noise diode
  to antenna temperature.  The right hand plot shows the individual
  beamswitched integrations averaged into 0.5 second measurements with
  error bars given by the internal scatter of each 0.5 second
  measurement and flux density calibrated via 3c286; here the initial
  10 second settle period has been excised. The remaining data
  illustrates the symmetric A/B/B/A nod pattern in which the source
  was placed in the two beams of the receiver.  The solid line in the
  right panel shows the template fit (Eq.~\ref{eq:gbtfit}).}
\label{fig:nod}
\end{figure}

For these observations we developed an ``On-the-Fly Nod'' variant of
the double-differencing technique described in \S~\ref{sec:ovroobs}
and \citet{readheadovro}. With this technique data are collected
continuously through the entire observation, including the slews
between beams. The recorded (10 Hz) antenna positions are used with
the target source coordinates to construct a template beamswitched
signal which is fitted to the differenced data.  This approach
minimizes scan-start overheads and provides a conveniently continuous
datastream for each source; it also allows us to carefully account for
imperfect source acquisition and settling times offline. An example
observation of a bright source is shown in Figure~\ref{fig:nod}. A
typical feature is the spike at $\sim 30 \sec$. In actuality the spike
is a dip in the source signal coming from the negative beam, and
arises from stiction in the telescope servo resulting in overshooting
the source slightly when slewing in one direction. Since this
overshoot was recorded by the antenna position encoders it is reflected
in the template model and has minimal effect on our observations,
particularly of our much weaker science sources. The on-source dwell
time in each phase was 10 seconds; there was a 10-second slew between A
and B phases. A 10 second settle was also allowed at the start of each
scan in order to allow possible GBT feedarm vibrations (occasionally
excited by the servo system at the start of a scan) to settle, but
this was never seen to be an issue in our frequent bright-source
calibration checks.  The average slew-time between program sources was
20 seconds, for an average total elapsed time per source measurement
of 90 seconds.  During the nod measurement the detected RF power in
each of 16 channels was recorded at 200 Hz by the CCB.

During daylight the GBT pointing and focus was checked every half hour
on a nearby bright source ($S_{31 \ghz} > 500 \mJy$); at night, this
was relaxed to every 45 minutes. During each pointing and focus check
a nod observation was also collected to monitor intraday stability of
our measurements. To monitor interday repeatability of our
calibration, we selected a steep-spectrum (NVSS/OVRO) source near each
CBI field as well and measured it at least once per observing session.

\subsection{GBT Data Reduction}
\label{sec:gbtreduc}

\begin{figure}[htbp]
\epsfxsize=6.2in
\epsffile{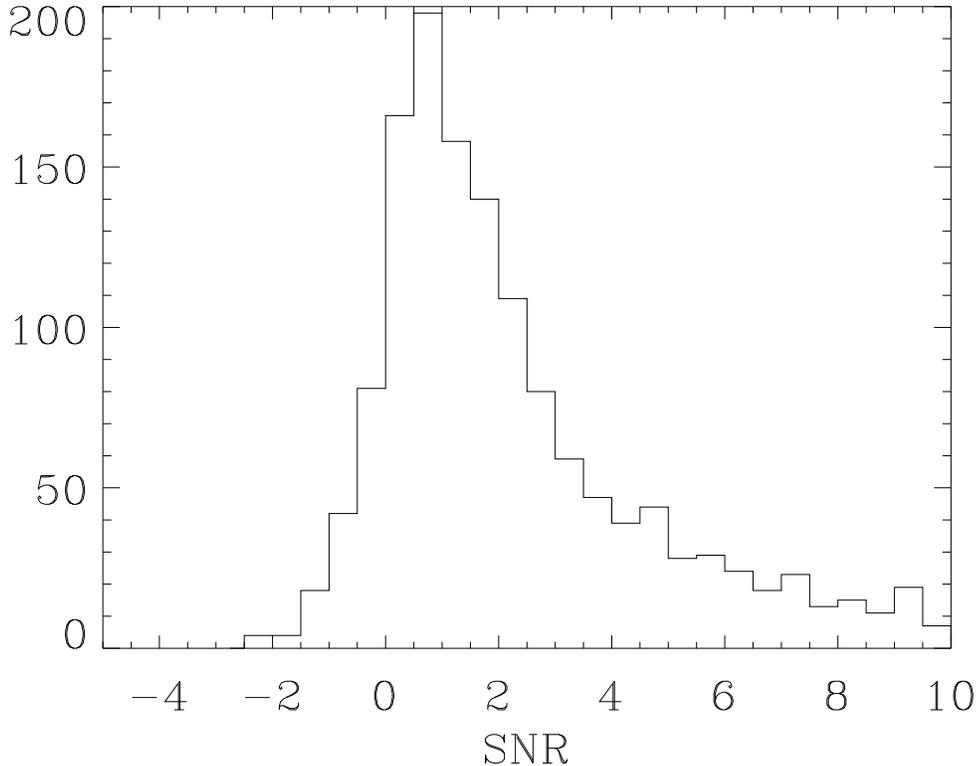}
\caption{The distribution of the signal-to-noise ratio for the
measurements in the GBT survey. Compare with blank-field
measurements in Fig.~\ref{fig:gbtnoisetest}.}
\label{fig:gbtsnr}
\end{figure}

While the GBT observations are similar to the OVRO observations,
several important differences led to a different approach to
reducing the data. First, substantially less ``data reduction''
(averaging) was performed by the online data acquisition system,
enabling preservation of more data for offline diagnostics. Second,
the typical signal-to-noise in a GBT observation was of order unity
owing to the higher sensitivity of the telescope -- see
Figure~\ref{fig:gbtsnr}. Finally the dataset was substantially smaller,
less than a week in total in contrast to several years of OVRO
observations. 

The CCB data were first averaged from 5 ms to $0.5 \, {\rm sec}$
integrations $d(t_i)$ and each integration was assigned an error
estimate based on the internal scatter of the beamswitched 5 millisecond
integrations. The 10 Hz recorded antenna positions were interpolated
onto the same time sampling as the CCB data resulting in a time series
of positions $\vec{x}_j(t_i) \equiv \vec{x}_{j,i}$ for a given feed
indexed by $j=1,2$ for a given observation.  The beam locations on the
sky and measured GBT beam pattern $B$ as a function of frequency were
used to compute the expected beamswitched response of the receiver to
a point source of flux density $s_o$ at the location of the source of
interest, $\vec{x}_o$
\begin{equation}
d_i = \, s_o  \times [ B( |\vec{x}_{1,i} - \vec{x}_o| ) - B( |\vec{x}_{2,i} - \vec{x}_o| )]
        + \langle d_i \rangle  + \frac{d d_i}{dt}
\label{eq:gbtfit}
\end{equation}
where the difference in square brackets comes about due to the beam
switching.  The last two terms are mean (radiometric offset) and
gradient terms allowed in the fit and which, due to the symmetry of
the nod pattern, are approximately orthogonal to the source flux
density parameter $s_o$.  This template was fit directly to the beamswitched
data -- refer again to Figure~\ref{fig:nod} for an example. The
$\chi_{\nu}^2$ of the fit is a good diagnostic of data quality; for
sources weaker than 10 mJy, $\chi_{\nu}^2$ was close to unity under
good conditions. As weather conditions degrade our simple model
fails resulting in appreciable increase of the $\chi_{\nu}^2$.  For
sources brighter than $\sim 10 \, {\rm mJy}$ $\chi_{\nu}^2$ rarely
approached unity even under excellent observing conditions due to
imperfections in the beam model and residual pointing errors. These
observations  required separate consideration in
$\chi_{\nu}^2$-based filters.

We performed this procedure separately on each of the 16 CCB data
channels.  Using observations of 3C286 throughout the observing
campaign a single mean calibration, referenced to the WMAP-5
Jupiter-temperature scale described in \S~\ref{sec:ovroobs}, was
determined for each channel. For the final processing this calibration
was applied to the individual detector timestreams and all timestreams
for a given frequency were averaged in the time domain before
performing the source flux density fit of Eq.~\ref{eq:gbtfit}. This
ensured that noise fluctuations correlated between feeds or
polarizations were correctly accounted for in the noise estimate that
follows.

Repeated observations of bright, steep-spectrum sources near our
fields (also of fainter steep-spectrum sources) provided a check on
the validity of our pointing filters.  We found that the accuracy of
our calibration for an individual source is 10\% (RMS) at 31 GHz,
dominated by uncertainties in the GBT pointing.

\begin{figure}[htbp]
\epsfxsize=6.2in
\epsffile{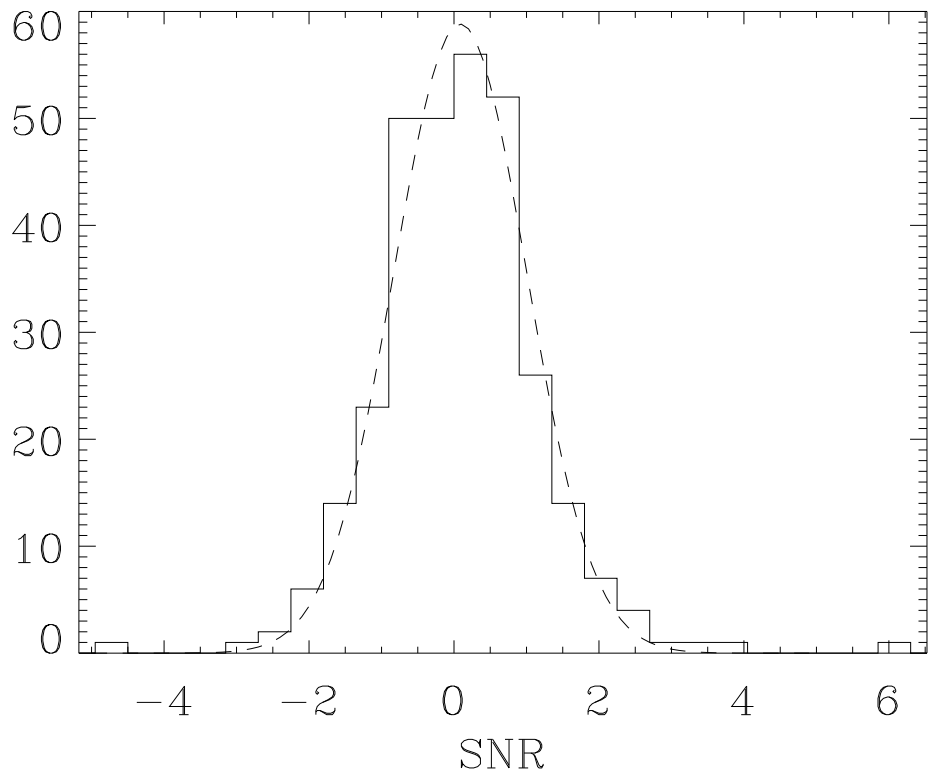}
\caption{Distribution of blank-field measurements divided by their
individual measurement noises estimated by
Eq.~\ref{eq:gbtnoiseest}. The dashed line shows the best fitting
Gaussian which has $\sigma = 0.91$, close to the expected value of
unity which would be obtained for perfectly estimated Gaussian noise;
the noise values themselves varied by a factor of 4. Compare to the
corresponding distribution of measurements that targeted NVSS sources
in the survey, Figure~\ref{fig:gbtsnr}.}
\label{fig:gbtnoisetest}
\end{figure}

Since the signal to noise in a typical 70-second GBT observation was of
order unity we could not straightforwardly compute the noise in our data
from the scatter of the measurements ({\it i.e.}, the fitted values of
the source flux densities $s_o$)  as we did for the OVRO
measurements.  Instead we formed alternate combinations of the
individual segments of the symmetric Nod procedure to quantify the
photometric stability of the measurement.  Designating the average
value of integrations within individual segments of the nod by
$A_1,B_1,B_2,A_2$, then $\Delta A = A_1-A_2$ is a source-signal-free
combination measuring to the photometric fluctuation over 50 seconds,
and $\Delta B = B_1-B_2$ is a source-signal-free combination measuring
the photometric fluctuation over 10 seconds. From test observations of
blank patches of sky under a wide range of conditions we found that the
average (over some window of many observations) of the
root-mean-squares of $\Delta A$ and $\Delta B$ gave an estimate of
the RMS of the measured flux density values $s_o$ accurate to within 10\%.
Note that since both $A$ and $B$
respond to time gradients of the beamswitched signal, whereas the
fitted source flux density ($s_o$ in Eq.~\ref{eq:gbtfit}) does not,
we might expect them to slightly overestimate the noise.
To improve the robustness of this
noise measure,  we computed the mean absolute deviation
(MnAD) of $\Delta A$ and $\Delta B$ rather than the RMS and
renormalize to a Gaussian equivalent ($MnAD = \sqrt{2/\pi} \sigma_g$);
we chose the somewhat less robust but lower variance MnAD over the
MedAD used in \S~\ref{sec:ovroreduc} because outliers are a lesser
concern for the signal-free estimators $\Delta A , \Delta B$ than in
the flux density measurements themselves.  For each measurement we
computed these quantities in a one-hour buffer centered on the
observation in question. With the appropriate normalization constants,
derived in the white-noise limit, this approach gave a noise estimate
\begin{equation}
N(t_i) = \frac{\sqrt{\pi}}{8} [MnAD(\Delta A) + MnAD(\Delta B) ]
\label{eq:gbtnoiseest}
\end{equation}
(where the mean absolute deviations were computed from all observations
within plus or minus a half hour of $t_i$) close to the RMS that
signal-free nod measurements would have given under the same conditions.
The data from the blank-field test observations are shown in
Figure~\ref{fig:gbtnoisetest}, normalized by the individual estimated
measurement errors. The actual per-measurement noises varied by a
factor of four. The dispersion in the noise-normalized data is $\sigma
= 0.91$, acceptably close to the $\sigma = 1.0$ which would result
from perfectly measured Gaussian noise. As expected the noise was
slightly overestimated due to the presence of radiometric
gradients. 

\begin{figure}[htbp]
\epsfxsize=6.2in
\epsffile{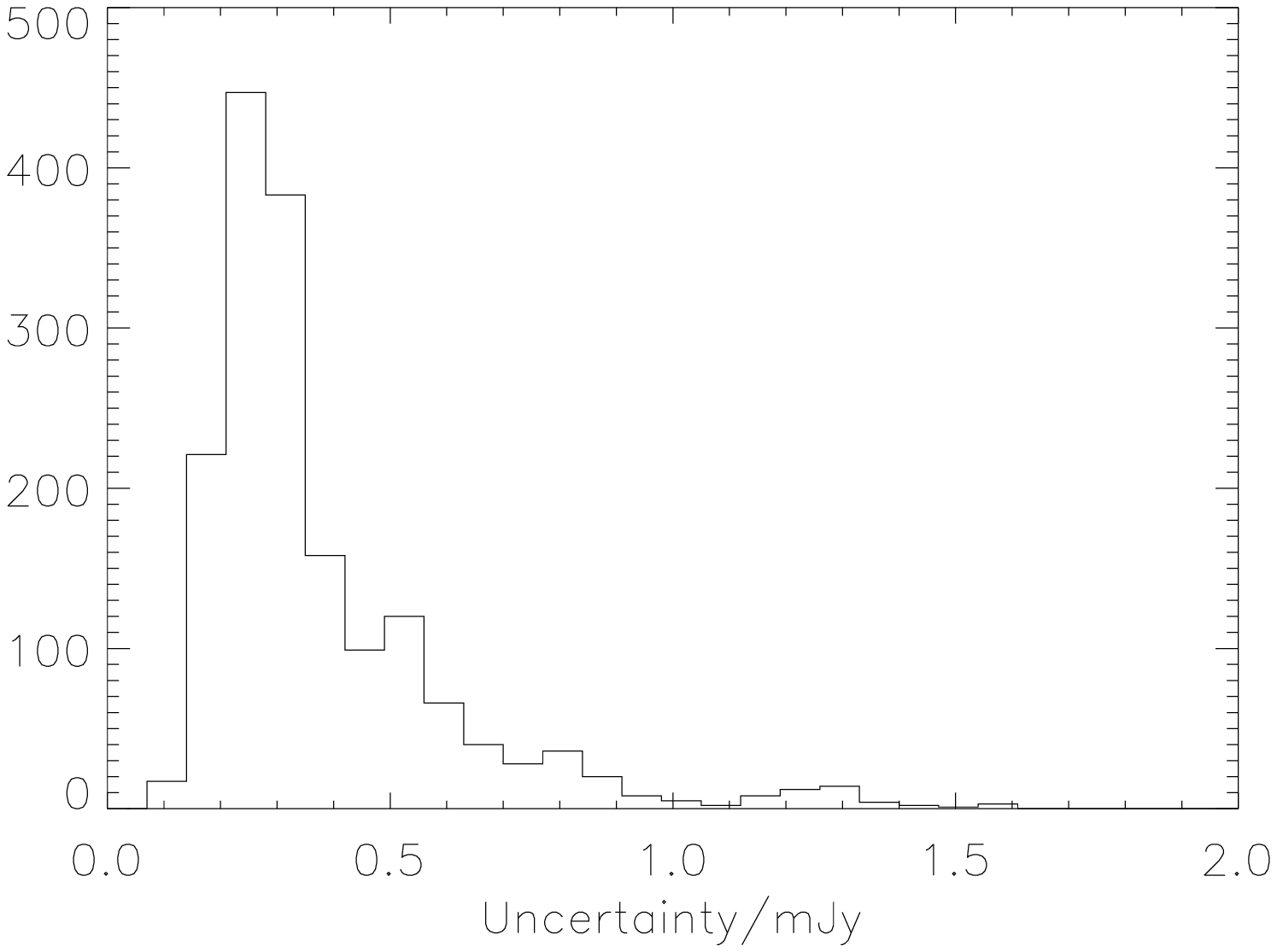}
\caption{Distribution of RMS measurement errors in the GBT survey,
estimated as described in the text.}
\label{fig:gbtnoisedist}
\end{figure}

The mean noise level in the GBT data which pass the data filters
described below is $390 \uJy$. The distribution is shown in
Figure~\ref{fig:gbtnoisedist}.

The main time-variable systematics that affect our data were
departures from good photometric conditions caused by atmospheric
clouds and water vapor and telescope pointing and focus. A series of
filters excised compromised observations. To identify periods of poor
photometry we computed the median $\chi_{\nu}^2$ in sliding one-hour
buffers -- excluding measurements brighter than 10 mJy -- and rejected
observations for which this value exceeds $1.4$. To catch rare
isolated glitches, individual observations with $\chi_{\nu}^2 > 1.4$
were also rejected {\it unless} the fitted flux density exceeded 10
mJy. In this case the observation was inspected manually; all passed
inspection so were retained.

In the course of commissioning and operations the pointing and focus
performance of the GBT was extensively characterized, with
results summarized by \citet{ptcs26} and \citet{ptcs49}. Under calm
conditions (winds under $3.0$ m/s) the observed RMS in the radial
pointing offset is $2.7''$, corresponding to less than 5\% loss of
peak signal for the GBT's $24''$ (fwhm) beam at 31 GHz.  When the
windspeed is greater than a few meters per second the GBT pointing
performance degrades, principally due to excitation of feedarm
vibrations; up to $4.0$ m/s, the pointing accuracy is still acceptable
for 31 GHz observations.  We rejected individual observations with mean
wind speeds over 3 meters per second; we also reject individual
observations with peak wind speeds over 5 meters per second. Every
half hour (during the day) to 45 minutes (at night), the telescope
pointing and focus was updated from observations of a nearby
calibrator.  Science program observations that  are
not preceeded by a successful peak and focus correction within these
periods of time are also rejected.  There were no GBT data for which
the sun or the moon were closer than 10 degrees away. The data filters
are summarized in Table~\ref{tbl:filters}.



To check the effectiveness of our pointing update criteria we
considered the ratio $r$ of measured flux density in the 38 GHz band
to that measured in the 31 GHz band for all sources detected at
$4\sigma$ or greater.  The flux density measured in the high frequency
band will fall more for a given pointing offset than will the lower
frequency band. For the dataset as a whole $r=0.84 \pm 0.03$, where
the uncertainty is the error in the mean assuming Gaussian statistics
from the RMS of the distribution. From the spectral index analysis of
\S~\ref{sec:maxlike} we calculated an expected $r=0.83 \pm 0.04$,
where the error bar in this case is the RMS of the distribution
(indicating its intrinsic width) predicted by our Maximum Likelihood
spectral index distribution under the assumption of a single power-law
extrapolation.  Note that selecting $4\sigma$ detections will bias $r$
slightly high relative to the full distribution.  Splitting the data
into two halves based on the time since the last pointing calibration,
the data with more recent corrections has an average flux density
ratio of $0.84$, while data with less-fresh pointing corrections have
an average flux density ratio of $0.83$, indicating no significant
change in the telescope pointing between pointing checks. Similarly
the daytime data have $r=0.83$ and the night-time data $r=0.84$,
indicating that on average the thermal effects prevalent during the
day do not significantly affect the telescope pointing.  The $\Delta r
= 0.01$ differences seen correspond to average radial pointing offsets
of $2''$ or $2\%$ gain effects; a 10\% loss in gain overall, if caused
by a pointing offset, would correspond to $\Delta r = 0.05$ (a $4''.6$
radial offset).  Note that this test is sensitive to any errors which
causes relative changes in observed power across the receiver
band. For instance, variations in the Ruze-equivalent RMS deviation of
the telescope surface from a perfect paraboloid scale as $exp(-(4\pi
\epsilon/\lambda)^2)$ and will also affect the $38$ GHz channel more
than the $31$ GHz channel.

\begin{table}
\begin{tabular}{|c||c|c|c|c|} \hline
Filter & Criterion & fraction of data passed \\ \hline
Wind & max $< 5.0 {\rm m/s}$, mean $< 3.0 {\rm m/s}$ & $79\%$ \\
$\langle \chi_{\nu}^2 \rangle$ & $< 1.4$ & 54\% \\
Pointing \& Focus Updates & within 30 (day) or 45 (night) minutes & $86\%$ \\ \hline
Total &  & 39\% \\
\hline
\end{tabular}
\caption{Summary of GBT data filters. We show the fraction of the
total dataset that passes each individual filter; since there are
correlations between the filtered variables the filters are not
statistically independent.}
\label{tbl:filters}
\end{table}

To further check the accuracy of our noise estimate we made use of the
fact that in the course of the survey 50 weak ($S_{31} < 15 \mJy$)
sources were observed more than once; we selected weak sources in order
that the effect of gain errors, which can be correlated between
observations, not be a dominant effect.  In all there are 122 such
observations. Subtracting the per-source means from each we calculate
a $\chi_{\nu}^2 = 0.95$ for $\nu = 72$. The probability to exceed this
by chance is 59\%.

\subsection{The Effect of Finite Source Size}

While most radio sources are compact compared to the GBT and OVRO
beams, a handfull are sufficiently extended that flux density is lost
in targetted observations.  On angular scales at which the GBT begins
to lose flux density ($\sim 10''$) most of the extended emission seen
in extragalactic sources originates in radio jets with a synchrotron
spectrum $\propto \nu^{-0.6}$ to $ \nu^{-1}$
\citep{laingpeacock80,dennetthorpe99}; the 31 GHz emission is
dominated by the compact, flatter-spectrum cores. Consider for example
that the NVSS at $1.4$ GHz, with a $45''$ (FWHM) beam find $20\%$ of
sources to be resolved, while 9C at 15 GHz, with a $25''$ FWHM beam,
finds only $9\%$ of sources to be detectably resolved
\citep{lizsizes}.  Consequently we expect that most of the flux
density in the 31 GHz sky will be in compact sources which are
accurately measured by the GBT.




To test for lost flux in the dataset NVSS sources with useful GBT
measurements were divided into two groups--- those resolved by NVSS
and those not resolved by NVSS--- and from these groups we constructed
two CBI visibility templates (actually, gridded estimator templates).
We fit for a scale factor $S_{CBI} = f \times S_{GBT}$ for each
template using the CBI visibility data in aggregate.  For sources
unresolved by NVSS we find $f = 0.96 \pm 0.04$, indicating that there
is no systematic bias between the flux density scales and that there
is no significant degree of source extension that is not detected by
NVSS.  Fitting for a CBI/GBT scale factor using only the NVSS-{\it
resolved} sources yields $f=1.18 \pm 0.10$.  The larger error bar in
this is consistent with the (smaller) number of extended sources in
comparison to the number of compact sources in the NVSS catalog.

We found that excising the extended sources from the spectral index
analysis of \S~\ref{sec:maxlike} did not significantly change the
spectral index probability density function (PDF) or final result for
the residual source correction. This is consistent with the expectation
that the 31 GHz sky variance is dominated by the flat, compact
sources.


\subsection{Confusion}

Since the angular resolution of the GBT and, especially, OVRO surveys
are comparatively low ($24''$ and $1'.3$, respectively) chance
superpositions of radio sources  occasionally occur and must be
considered.  To assess the effects of source confusion we combined the
OVRO and GBT catalogs, giving precedence to GBT measurements where
present, and selected $3\sigma$ or greater detections. To this we
added the NVSS sources with $1.4 \, {\rm GHz}$ flux densities $> 3.4
\mJy$ and multiplied their flux densities by $0.1$ (the mean ratio of
$S_{31}$ to $S_{1.4}$ determined from the Maximum Likelihood analysis
of \S~\ref{sec:maxlike}), thereby obtaining our best estimate of the
point sources 31 GHz sky in the regions observed. We call this our
``reference'' catalog.

Using the reference catalog we scanned the full set of OVRO observations
and identified those for which the sum of the absolute values of
beam-weighted confusing source flux densities in the reference catalog
amounts to more than half the measurement error for that source.  This
amounts to $1.3\%$ of the OVRO catalog. For these measurements a
correction is calculated using the procedure described in
Appendix~\ref{appendix:confusion}.

Owing to the smaller beam size and beam throw the level of source
confusion in the GBT data was much lower.  Using the same criteria
only two observations were significantly confused; for these sources
the correction described in Appendix~\ref{appendix:confusion} was also
performed.

\subsection{Source Catalogs}
\label{subsec:catalogs}

The full set of OVRO observations is presented in
Table~\ref{tbl:ovroresults}, and the GBT survey results are presented
in Table~\ref{tbl:gbtresults}.  Reported error bars include a 10\% and
5\% RMS gain uncertainty for GBT and OVRO measurements,
respectively. Sources detected at greater than $4\sigma$ at 31 GHz are
marked; for this calculation the random gain uncertainty is excluded.
In all $3,165$ sources were observed.  The GBT catalog contains
$1,490$ sources.  Of the $2,315$ useful OVRO observations many of the
non-detections (and a few detections) are superceded by more sensitive
GBT observations; the OVRO catalog therefore contains data on $1,675$
sources. The detection rate of the OVRO measurements was $11\%$, and
that of the GBT measurements $25\%$. In all $18\%$ of sources were
detected at 31 GHz.

Also included in the table are the $1.4$ GHz flux densities, source
sizes from the NVSS catalog, flags to indicate $4\sigma$ detections,
and flags to indicate which observations have been corrected for the
effects of source confusion in either the main or reference beams.

The catalogs presented here are based on the processing described in
\S~\ref{sec:ovroreduc} and \ref{sec:gbtreduc}. The analysis
\S~\ref{subsec:srccounts} and \S~\ref{subsec:cbipow} used an earlier
processing with slightly less strict filters. The total number of
sources in this catalog was $3,562$. The spectral index distributions
obtained from these two versions of the source lists are consistent.

\begin{table}
\begin{verbatim}
 Name         RA/J2000    Dec/J2000    S30   E(S30) S(1.4)  E(S1.4)Maj   Min  D C 
085057-0150 08 50 57.06 -01 50 38.6    5.30  1.91    46.90   1.50   0.0   0.0
085101-0509 08 51 01.80 -05 09 52.9    3.00  3.40    36.30   1.20   0.0   0.0
085103-0303 08 51 03.94 -03 03 35.5    2.00  2.20    17.80   1.40  71.4   0.0
085118-0419 08 51 18.93 -04 19 10.3   -3.13  2.81    62.70   2.30  15.3   0.0   *
085121-0418 08 51 21.49 -04 18 25.7   11.74  2.79    66.00   2.40  45.5   0.0 * *
085127-0156 08 51 27.01 -01 56 09.3    4.00  5.00    18.70   1.00  22.7   0.0
085130-0155 08 51 30.58 -01 55 49.4   -4.30  3.80    22.70   0.80   0.0   0.0
085135-0150 08 51 35.60 -01 50 44.9    5.80  2.71    55.20   1.70   0.0   0.0
085137-0405 08 51 37.67 -04 05 00.7    3.40  3.30    27.40   0.90   0.0   0.0
085138-0451 08 51 38.97 -04 51 23.8   14.20  3.25    78.40   2.40   0.0   0.0 *
085141-0424 08 51 41.73 -04 24 35.6    2.80  2.10    13.70   0.60   0.0   0.0
085149-0314 08 51 49.11 -03 14 57.0    0.70  2.50    81.40   2.90  29.8   0.0
085157-0408 08 51 57.40 -04 08 01.2   -0.10  1.80    17.30   0.70   0.0   0.0
\end{verbatim}
\caption{Excerpt of OVRO 40-m survey results. Positions and $1.4$ GHz
flux densitiesare from NVSS. Columns are: NVSS name; Right Ascencion
(J2000); Declination (J2000); 31 GHz flux density and uncertainty in mJy;
NVSS integrated flux density and uncertainty; NVSS Major axis in
arcseconds ($0.0$ indicates no detected size); NVSS minor axis in
arcseconds. A flag in the ``D'' column indicates a $4\sigma$
detection, and a flag in the ``C'' column indicates that a confusion
correction has been performed by the method described in the text. The
full version of this table is available on line.}
\label{tbl:ovroresults}
\end{table}

\begin{table}
\begin{verbatim}
 Name         RA/J2000    Dec/J2000    S30   E(S30) S(1.4)  E(S1.4)Maj   Min  D C 
024033-0430 02 40 33.46 -04 30 00.5    0.17  0.30     3.60   0.60   0.0   0.0
024033-0432 02 40 33.53 -04 32 47.5    0.37  0.32     5.50   0.50   0.0   0.0
024038-0425 02 40 38.89 -04 25 54.5    0.81  0.32     4.00   0.50   0.0   0.0
024055-0428 02 40 55.18 -04 28 36.4    0.34  0.30     4.50   0.50   0.0   0.0
024108-0422 02 41 08.83 -04 22 51.4    5.32  0.61     6.90   0.50   0.0   0.0 *
024111-0425 02 41 11.61 -04 25 21.4    1.23  0.32    26.80   0.90   0.0   0.0 *
024119-0421 02 41 19.39 -04 21 44.4    2.71  0.43    17.20   0.70   0.0   0.0 *
024129-0003 02 41 29.53 -00 03 27.4   -2.40  1.19     4.70   0.50   0.0   0.0
024137-0039 02 41 37.85 -00 39 19.4    0.56  0.61    14.60   0.60   0.0   0.0
024144-0416 02 41 44.17 -04 16 48.0    5.88  0.67    84.50   3.30  32.6  16.5 *
024146-0025 02 41 46.15 -00 25 01.7    0.89  0.61     5.70   0.50   0.0   0.0
024153-0105 02 41 53.70 -01 05 43.3    0.15  0.24     3.40   0.50   0.0   0.0
024204-0053 02 42 04.56 -00 53 33.7    0.26  0.24     5.60   0.50   0.0   0.0
\end{verbatim}
\caption{Excerpt of GBT 31 GHz survey results. Columns are as in Table~\ref{tbl:ovroresults}.}
\label{tbl:gbtresults}
\end{table}

\section{Interpretation}
\label{sec:results}

\subsection{Spectral Index Distribution from GBT, OVRO, and NVSS Data}
\label{sec:maxlike}

In order to determine the contribution of radio sources below the NVSS
completeness limit to the sky variance measured by 31 GHz CMB
experiments such as the CBI, it is necessary to understand the $1.4$
to $31$ GHz spectral index distribution --- or equivalently, the
probability density function (PDF) of $S_{31}$ given $S_{1.4}$. The
data from the survey presented in this paper are the best currently
available for this purpose. Since sources detected at 31 GHz are preferentially flat-spectrum, it is
necessary to include non-detections in this analysis to obtain an
unbiased result. Because we know the low-frequency flux densities of
these sources and the 31 GHz measurement noise, these non-detected
sources will impose constraints on the spectral index distribution.
We adopted a Bayesian Maximum Likelihood approach.

We wish to find the spectral index distribution that maximizes the
likehood of measuring the observed 31 GHz flux densities given their observed
1.4 GHz NVSS flux densities.
The general form of the likelihood of measuring 31 GHz flux density $\skaobs$
given 1.4 GHz flux density $\slobs$ is:
\begin{equation}
P\left(\skaobs | \slobs \right) = \int \int P\left(\skaobs | \skatrue \right)
P\left(\skatrue | \sltrue\right) P \left(\sltrue | \slobs \right) d\skatrue d\sltrue
\end{equation}
integrating over the unkown values of the ``True'' fluxe densities 
$\skatrue$ and $\sltrue$, and with the
(unknown) 1.4-31 GHz spectral index function $ P\left( \skatrue |
\sltrue \right) $.  We parameterized $\sratio$ with a set of
$N_{bin}=17$ points in the frequency spectral index $\alpha$ evenly
spaced between $\alpha=-1.6$ and $\alpha=+1$. Appendix~\ref{appendix:maxlike}
contains a more detailed discussion of the evaluation of this
likelihood function.

To measure the the PDF and its uncertainty, we used the publicly
available Markov-Chain Monte-Carlo (MCMC) code COSMOMC \citep{cosmomc}
adapted for use with a generic likelihood function.  The MCMC
algorithm draws samples a multi-dimensional parameter space, in this
case the space of $N_{bin}$ parameters representing the spectral index
distribution, with a specified distribution, in this case the
likelihood of the parameters given the data.  This procedure permits
easy evaluation of the uncertainties and covariances in parameters of
interest (the spectral index distribution); it also makes evaluating
the uncertainties and covariances in functions of these parameters
straightforward ({\it e.g.}, \S~\ref{subsec:srccounts}).

\begin{figure}
\centering
\includegraphics[width=16cm]{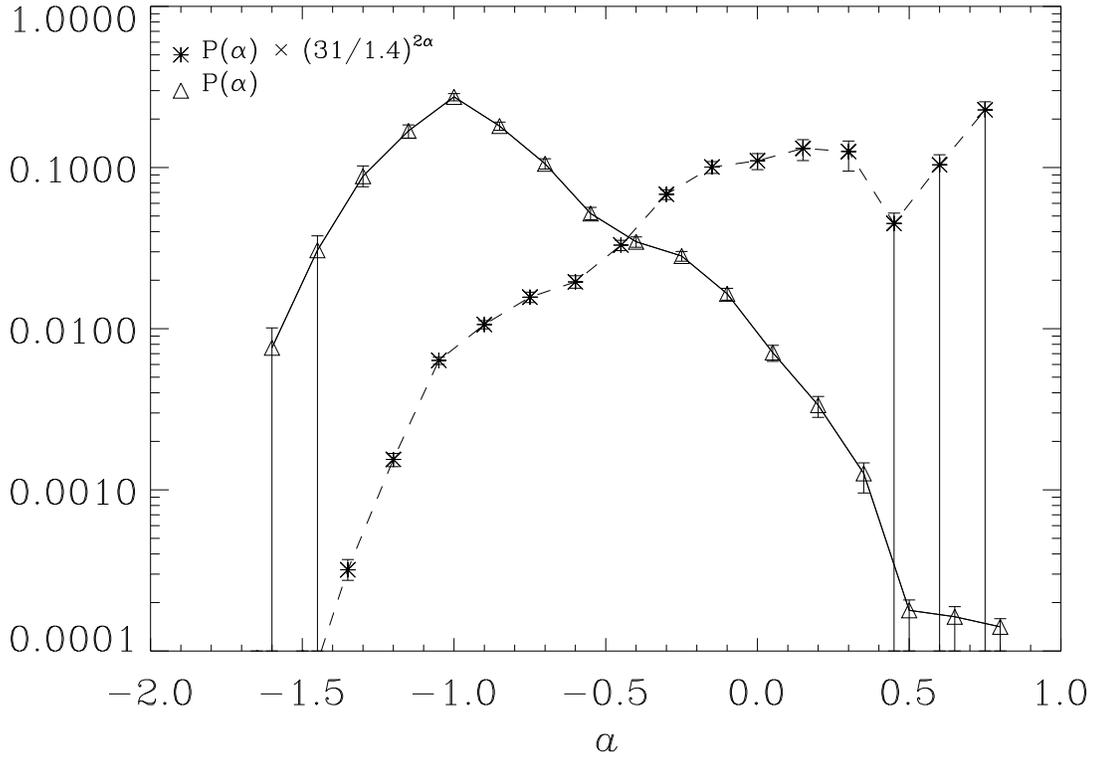}
\caption{The $1.4$ to 31 GHz spectral index distribution determined from
GBT, OVRO, and NVSS data.  Error bars are the 1-parameter marginalized
$1\sigma$ uncertainties for the individual points parameterizing the
spectral index PDF. Also shown is the PDF weighted by $(31/1.4)^{2\alpha}$,
which is proportional to the contribution that sources of a given
$\alpha$ make to the variance of the 31 GHz sky intensity.}
\label{fig:specind}
\end{figure}

We show the marginalized posterior distributions of the parameters of
the spectral index PDF in Figure~\ref{fig:specind}.  This is our best
description of $1.4$ to 31 GHz source spectral indices. This figure
also shows the spectral index distribution weighted by $(31/1.4)^{2
  \alpha}$, and therefore proportional to the variance in 31 sky
brightness that residual point sources of a given spectral index
contribute.  Table~\ref{tbl:specind} summarizes the results in binned
form, calculated from $100,000$ samples from the MCMC chains.  We
found that the mean $31$ to $1.4$ GHz flux density ratio is $0.111 \pm
0.003$, corresponding to a spectral index $-0.71 \pm 0.01$. The
distribution is heavily skewed towards steep spectral indices with a
long tail in the flat direction, resulting in a {\it mean spectral
  index} steeper than the value $\alpha=-0.71$ that corresponds to the
mean flux density ratio ($<\alpha>= -0.92^{+0.29}_{-0.30}$, $68.5 \%$ confidence
interval).  $9.0 \pm 0.8\%$ of sources have spectral indices flatter
than $\alpha > -0.5$ and $1.2 \pm 0.2\%$ have inverted spectral
indices, $\alpha > 0$.

\begin{figure}
\centering
\includegraphics[width=16cm]{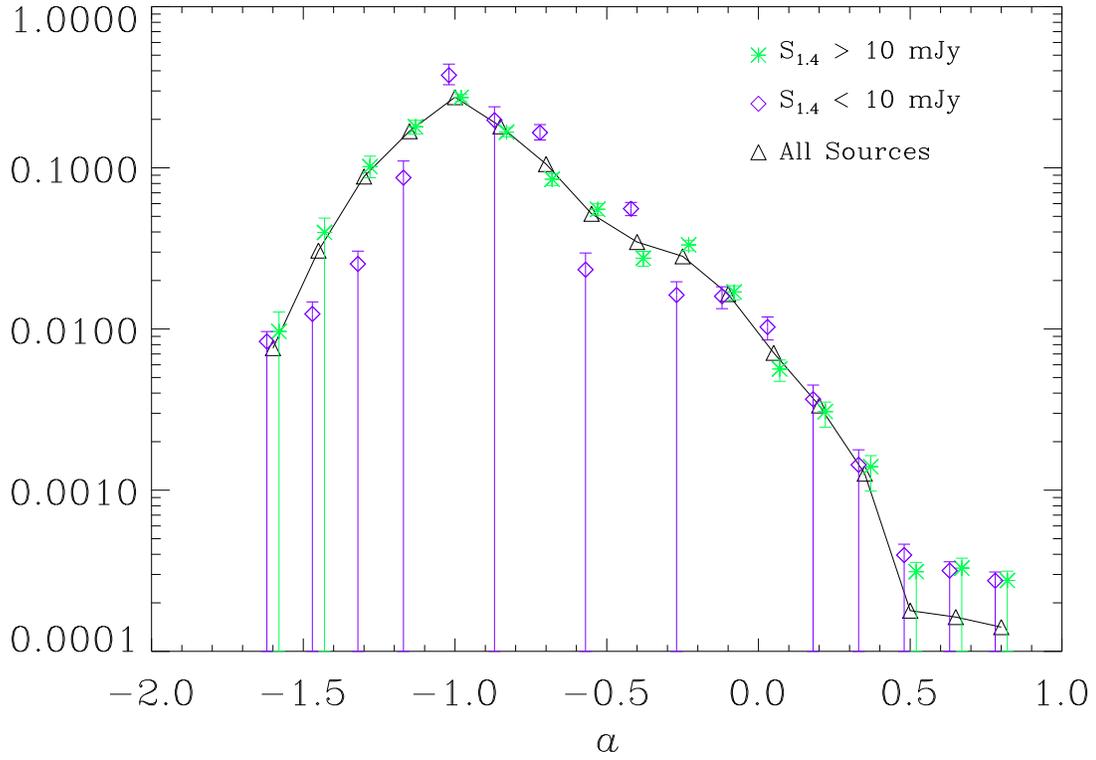}
\caption{The spectral index from $1.4$ to 31 GHz determined from
bright ($S_{1.4} > 10 \mJy$) and faint ($S_{1.4} < 10 \mJy$)
subsets of the full dataset. We also show the spectral index
distribution from the full dataset.}
\label{fig:specindsplit}
\end{figure}

To check the important assumption that the spectral index does not
change as a function of $1.4$ GHz flux density we split the sample
into $S_{1.4} > 10 \mJy$ and $S_{1.4} < 10 \mJy$ subsamples and
estimated the spectral index distribution for each of the bright and
faint samples separately, with results shown in
Figure~\ref{fig:specindsplit}. At low flux densities the faint
subsample provided little constraint on the spectrally steep end of
the distribution, reflected by large error bars in that regime. The
consistency of the subsamples supports the assumption that the
spectral index distribution is constant over the range of flux
densities of interest. To further assess the robustness of our
conclusions we re-ran the spectral index distribution chains varying
the assumed noise level of the OVRO and GBT data by $\pm 20\%$ and
excised potentially confused sources. The results, along with the
nominal case, are summarized in Table~\ref{tbl:sitests}. We show the
mean spectral index and its RMS, the fraction of sources with flat our
rising spectra, the mean 31 to $1.4$ GHz flux density ratio, and the
mean-square flux density ratio, which is more indicative of the
variance of the source population.


\begin{table}
\begin{tabular}{|c|c|c|c|} \hline
Test:                                  & Noise$\times 0.8$     & Nominal &  Noise $\times 1.2$  \\ \hline
$f_{\alpha \ge 0.0}$                    & $1.24 \pm 0.15 \%$   &  $1.17 \pm 0.15 \%$ &  $0.92 \pm 0.18 \%$   \\
$\langle S_{31}/S_{1.4}\rangle$         &  $0.117$             & $0.111$             & $0.101$    \\
$\langle (S_{31}/S_{1.4})^2 \rangle$    &  $0.099$             & $0.092$             & $0.084$   \\
$\langle \alpha\rangle$                 & $-0.911$             & $-0.917$            & $-0.925$    \\
$\langle\sigma_{\alpha}\rangle$         & $0.336$              & $0.311$             & $0.292$   \\
\hline
\end{tabular}

\caption{Results of tests of the spectral index distribution
estimate. We show the fraction of rising spectrum sources, the mean
flux density ratio ($1.4$ to $31$ GHz), the mean of the square of the
flux density ratio (which is directly relevant to the residual source
variance), and the mean spectral index for a range of perturbed GBT
and OVRO noise levels, and for the case where all measurements
potentially affected by confusion are excised.}
\label{tbl:sitests}
\end{table}

\begin{table}
\begin{center}
\begin{tabular}{|c|c|} \hline
Spectral Index & $f/\%$ \\ \hline
   -1.60   &        $<1.8$      \\
   -1.45	  &  $3.1 \pm 1.6$          \\
   -1.30	  &      $8.8 \pm 2.8$     \\
   -1.15	  &       $16.9 \pm 3.3$  \\
   -1.0	  &       $27.4 \pm 3.0$      \\
   -0.85	  &      $18.1 \pm 2.4$       \\
   -0.7	  &      $10.6 \pm 1.6$       \\
   -0.55	  &      $5.2 \pm 1.0$      \\
   -0.40	  &      $3.5 \pm 0.6$      \\
   -0.25	  &      $2.8 \pm 0.4$       \\
   -0.10	  &      $1.6 \pm 0.3$       \\
   0.05	  &      $0.7 \pm 0.2$       \\
   0.20	  &      $0.3 \pm 0.1$       \\
   0.35	  &      $0.1 \pm 0.1$       \\
   0.50	  &      $<0.05$      \\
   0.65	  &      $<0.05$       \\ 
   0.80	  &      $<0.04$       \\
%
%
\hline 
\end{tabular} 
\end{center} 
\caption{ The $1.4$ to $31$ GHz spectral index distribution determined
  from GBT, OVRO and NVSS data. For bins consistent with zero at
  $1\sigma$, $2\sigma$ upper limits computed from the Likelihood
  function are listed.}
\label{tbl:specind}
\end{table}

\subsection{31 GHz Source Counts}
\label{subsec:srccounts}

\begin{figure}[htbp]
\epsfxsize=6.2in
\epsffile{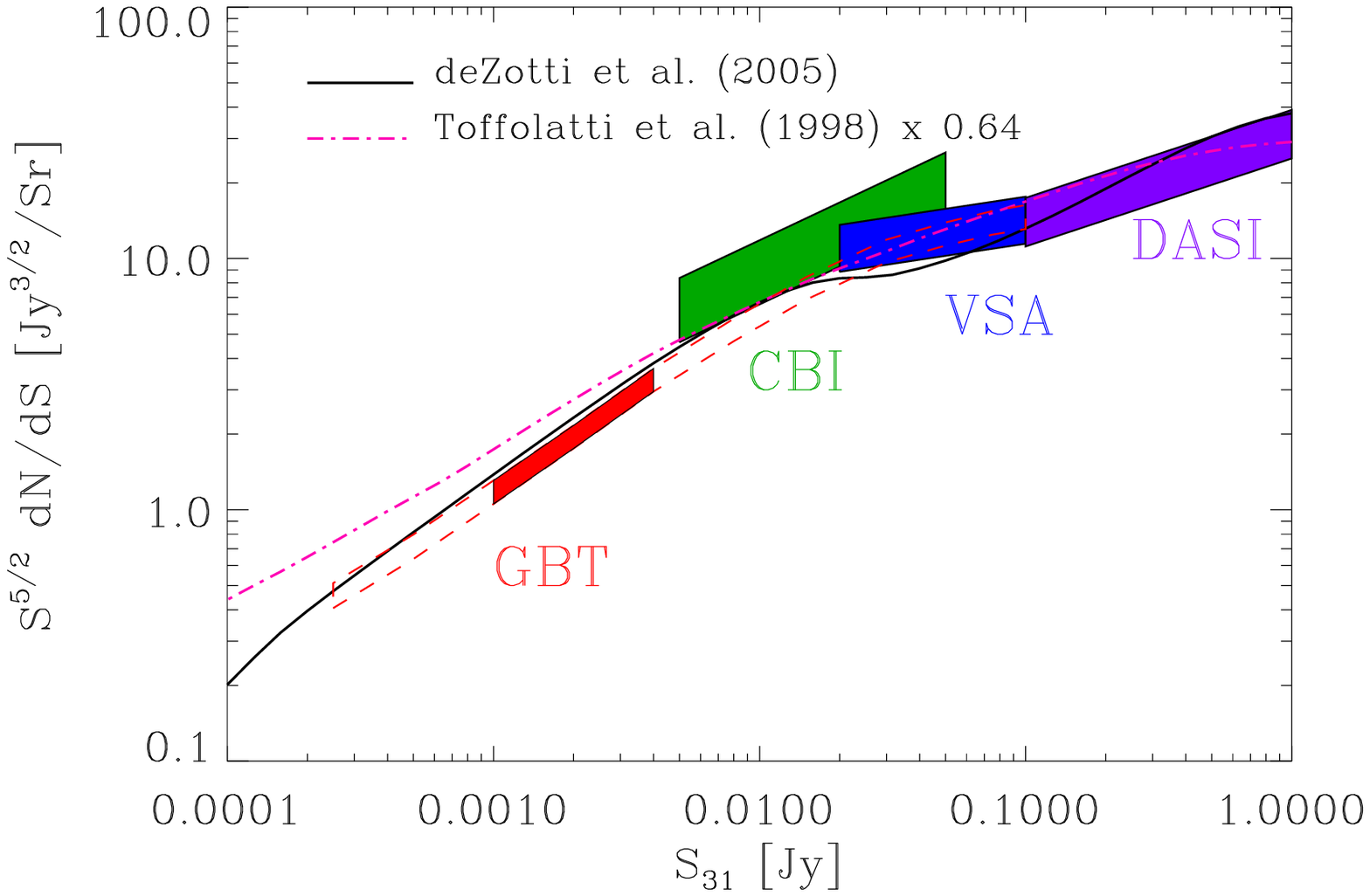}
\caption{A summary of 31 GHz source count measurements and models. CBI results are
from \citet{cbideep}, VSA results from \citet{cleary05}, and DASI
results from \citet{kovac02}. All data are at 31 GHz except for VSA,
which is at 33 GHz. Also shown is the \citet{toffolatti98} 31 GHz
model scaled down by a factor of $0.64$ and the \citet{dezotti05} 33
GHz model. The derivation of the GBT error box is described in the
text; the solid red box shows our best estimate of the counts, valid
over $1 \mJy < S_{30} < 4 \mJy$ and the red dashed line shows the
result over the full range $0.25 \mJy < S_{30} < 100 \mJy$.  Other
experiments' errors are taken from the Poisson error in the count
normalization.}
\label{fig:counts}
\end{figure}

We estimated the 31 GHz counts by by drawing flux densities
between $50 \, {\rm \mu Jy}$ and $1 \, {\rm Jy}$ from the Hopkins et
al. $1.4$ GHz counts and extrapolating them to 31 GHz with samples
from our $1.4$ to $31$ GHz PDF.  Over the range $1 \, {\rm mJy} <
S_{31} < 15 \, {\rm mJy}$ the counts follow a power law $dN/dS \propto
S_{31}^{-1.8}$, with a normalization at 1 mJy of $S^{5/2} \, dN/dS = 
1.18 \pm 0.05 \, {\rm Jy^{-1.5} \, sr^{-1}}$. Directly summing the
simulated source populations gives an integrated source count of
\begin{equation}
N(>S_{31}) = (16.7 \pm 1.7) \times (S_{31}/{\rm 1 mJy})^{-0.80 \pm 0.07} \, {\rm deg^{-2}}.
\end{equation}
As shown in Figure~\ref{fig:counts} our 31 GHz counts compare
favorably with other measurements \citep{cbideep,cleary05,kovac02}.
We can apply this procedure over an arbitrary range of flux densities
but it is only valid over the range that our spectral index PDF is
valid.  Based on the observed $31$ to $1.4$ GHz PDF we estimate that
the potentially distinct source populations at $S_{1.4} < 1 \mJy$ and
$S_{1.4} > 100 \mJy$ could contribute 10\% or more of the sources at
$S_{31} < 1 \mJy$ and $S_{31} > 4 \mJy$, respectively, so take this to
define the range over which our counts are valid. These potential contributions
to the counts also set the uncertainty in the power law slope. Error bars were
checked using the set of MCMC-sampled spectral-index PDFs, but are
dominated by the assumed 10\% systematic uncertainty. However the
agreement with models and data is good over a much wider range of flux
densities, suggesting that the change of the source spectral indices
is not especially strong.  The GBT counts have a similar slope to the
model of \citet{dezotti05}, although the normalization of the GBT
counts is $15\%$ lower in the $1-10$ mJy range. The
\citet{toffolatti98} counts are substantially higher than both in this
range. Below $0.3 \, {\rm mJy}$ the 31 GHz counts in our model show a
weak turn-up due to the sub-mJy population but the precise location
and magnitude of this turn-up depends on the assumed spectra of these
sources (see \S~\ref{subsec:otherpops}).

\subsection{The Effect of Unidentified Sources on CBI Measurements}
\label{subsec:cbipow}

\subsubsection{Simulations}
\label{subsec:cbisims}

Point sources are the largest astrophysical foreground in the CBI
data, and are especially critical at high-$\ell$.  An accompanying
paper \citep{cbi10} presents the power spectrum from 5 years of CBI
observations.  We have used the results of GBT and OVRO 40-m
measurements presented here to quantify the impact of discrete sources
on the power spectrum.  As discussed in \S~\ref{sec:intro} there are
two distinct classes of sources to be treated: those which are
individually known and identified from imaging surveys (NVSS); and
sources in our fields not detected in any survey, but expected to be
present based on source counts extending below the survey detection
limits in other fields.  Very conservatively, all known sources are
projected out of the CBI dataset; the efficacy of this procedure is
quantified in \citet{cbi10}. Our task, and the fundamental aim of this
paper, is to quantify the statistical contribution of the fainter
sources. This population is a {\it low-frequency selected population}
(sources below the NVSS detection threshold), and we must calculate
the variance of its sky brightness at 31 GHz.

\begin{figure}
\centering
\includegraphics[width=16cm]{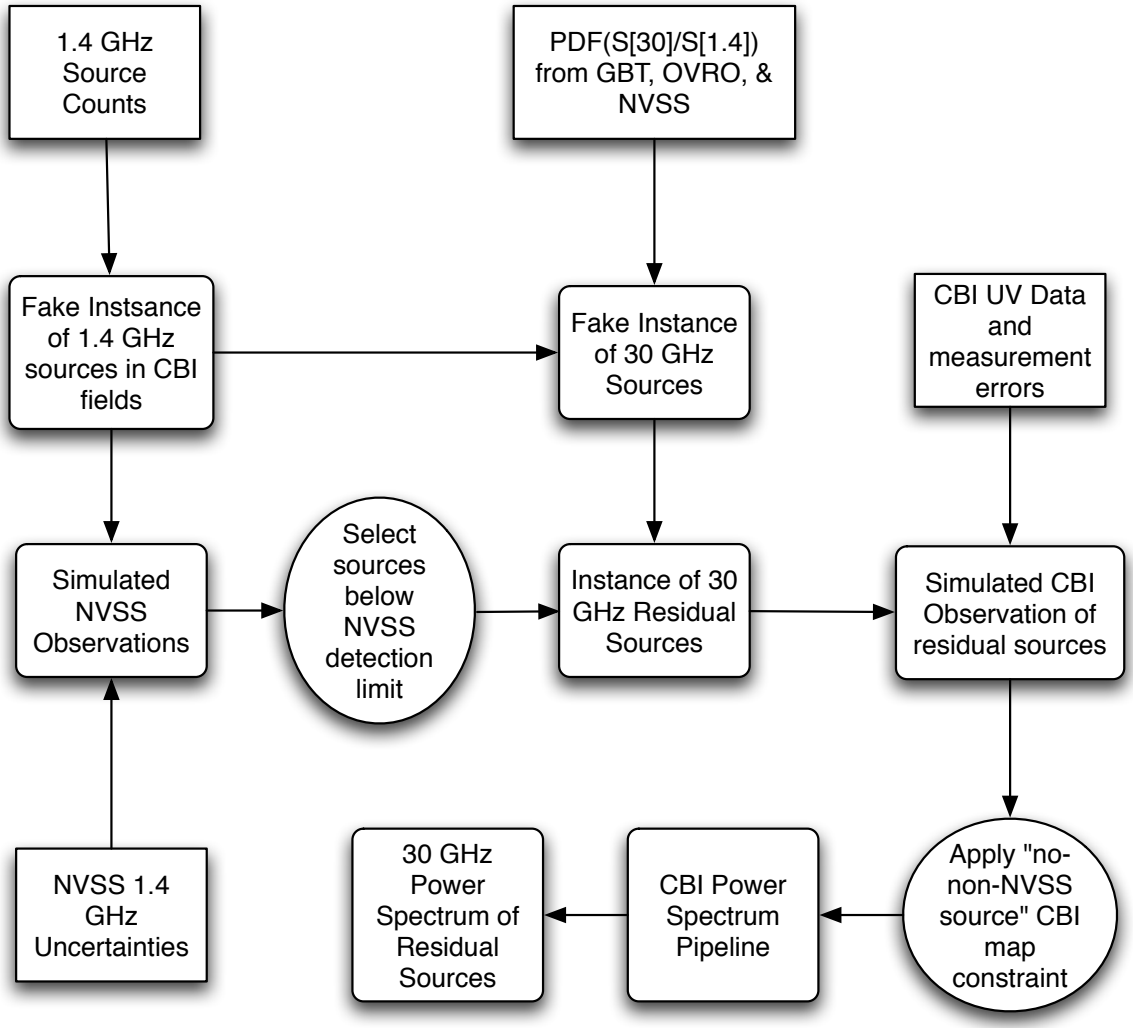}
\caption{Schematic of the simulation pipeline we use to estimate the
   distribution of residual source power in the CBI power spectrum.}
\label{fig:sourceplan}
\end{figure}

To do this we undertook an extensive suite of simulations. We created
realistically constrained realizations of the sub-NVSS populations and
ran these realizations through the full CBI power spectrum pipeline;
the procedure is schematically illustrated in
Figure~\ref{fig:sourceplan}.  We first drew $1.4$ GHz populations down
to 0.2 mJy using a power-law fit to the FIRST counts \citep{white97}
between 2 and 100 mJy, representing the dominant contribution of
mJy-level AGN.  The contribution of sources below 1 mJy at $1.4$ GHz,
which likely have different spectral indices than the mJy AGN, is
considered separately in \S~\ref{subsec:otherpops}.  We then simulated
NVSS observations of these source realizations, adding Gaussian noise
typical of the NVSS thermal noise ($0.6$ mJy).  Any source that has an
observed (noisy) flux density greater than our NVSS projection
threshold of $3.4$ mJy was then removed, leaving a realistic
population of $1.4$ GHz sources that would not have appeared in NVSS.
Sources on the power-law $dN/dS$ with $S_{1.4} < 0.2 \mJy$, which we
do not simulate, will contribute $< 2\%$ of the power.  We then drew
one $1.4$-$31$ GHz spectral index distribution from the Markov chains
in \S~\ref{sec:maxlike}, with the assumption that the spectral index
distribution is independent of $1.4$ GHz flux density, and assigned
each source a spectral index drawn from the distribution.  The signal
from these faint-source realizations was added to CMB+noise
simulations of the CBI dataset.

An additional constraint came from the fact that the CBI maps, at a
typical $5\sigma$ level of $S_{31} = 20 \mJy$, shows no sources that
are not present in NVSS. This limits the strongly inverted-spectrum
tail of the spectral index distribution.  At the NVSS lower flux limit
roughly one in three CBI synthesized beams ($\sim 5'$ FWHM) has an
NVSS source in it, so at the fainter $1.4$ GHz flux densities
characteristic of the residual sources chance superpositions 
will be common. Such a blend would appear to the CBI as a single
orphan source.  Therefore the absence of non-NVSS sources in the 31
GHz CBI maps also constrained the abundance of more modestly inverted
spectrum sources {\it for the particular realization of sources
  present in the CBI fields}. In this analysis we found that the
latter was the more important constraint.

To fold in this constraint we imaged each simulation and searched it
for ``orphan'' sources using the method described in \citet{cbi12},
and rejected any realization where such a source is found.  There were
a 4 $\sim 5-\sigma$ features in the CBI maps that were marginally
inconsistent with being associatable with an NVSS source(s).  We
followed each of these up with the GBT and do not detect them,
implying that they are probably noise fluctuations.  It is still
possible that these fluctuations were chance superpositions of
multiple faint sources in the same CBI beam.  The GBT could miss such
a superposition if no individual source falls within the GBT's
much-smaller beam when pointed at the effective emission center.  To
account for this possibility, we carried out mock GBT observations on
fluctuations in the simulated maps that were classified as orphan
sources, and do \textit{not} reject any simulation based on map
fluctuations that would not have been seen by the GBT. This gave us a
set of simulated source catalogs that are consistent with the observed
$1.4$-$31$ GHz spectral index distribution and the fact that CBI
detects no orphan 31 GHz sources.

We created 500 simulations, 250 for each of the binnings in
\S~\ref{sec:maxlike}, and subjected them to the map test described
above.  In 215 of the 500 (the ``clean'' simulations), no orphan
sources are detected in the simulated CBI maps.  In 285 of them (the
``dirty'' simulations) one or more orphan source is detected.  We take
the visibilities from the source simulations and run them through the
full CBI power spectrum pipeline, fitting an $\ell^2$ model to
determine the residual source power spectrum. 

\begin{figure}
\centering
\includegraphics[width=16cm]{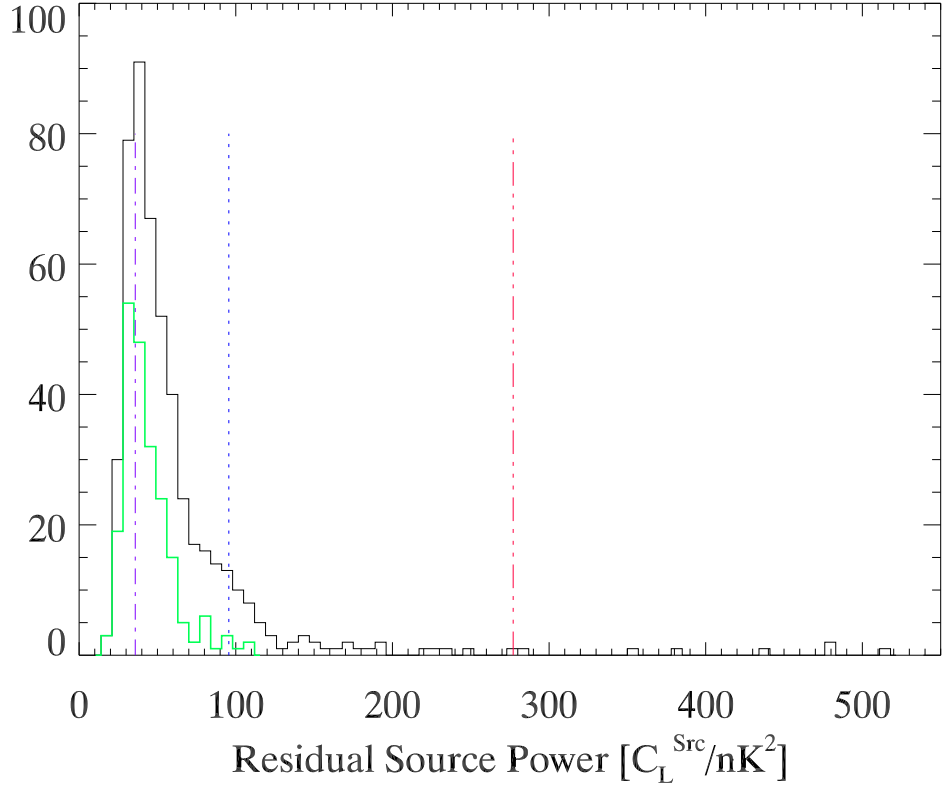}
\caption{The CBI residual source contamination, determined from
low-frequency source counts plus GBT and OVRO 31 GHz information, via
simulations described in the text. The heavy black line is the
distribution full distribution and is our best prediction for random
$\sim 140 \, {\rm deg^2}$ of sky.  The light green distribution is for
realizations which have no ``orphan'' sources which would have been
detected in the CBI maps, and represents our best prediction for the
residual source contribution to the CBI data.  The residual source
correction used in CBI analyses prior to this work is shown as a
dotted blue line; the dash-dot purple line is that calculated by
\citet{cleary05} from VSA source counts at 31 GHz; and the red
dash-triple-dot line is the level of source contamination needed to
fully account for the power that CBI observes in excess of intrinsic
CMB anisotropy \citep{cbi10}. Note that the units of the x-axis are $C_{\ell}$
rather than $\ell (\ell +1) C_{\ell}/2\pi$.}
\label{fig:isohist}
\end{figure}

The resulting estimate of the CBI residual source contamination is
shown in Figure~\ref{fig:isohist}.  We find the mean
signal\footnote{Previous CBI analyses expressed the residual source
  correction in the units implied by Eq.~\ref{eq:csrc}, while here we
  express them in terms of $C_{\ell}$, which is also independent of
  $\ell$ for unresolved sources but is more readily compared to other
  experiments. The conversion between $X_{src}$ and $C_{\ell}^{src}$
  is given by $C_{\ell}^{src} = X_{src} \times \left[\frac{2 k_B}{c^2}
    \left(\frac{k_B T_{cmb}}{h}\right)^2 \, \frac{x^4 e^x}{(e^x-1)^2}
    \right]^{-2}$. Note that CMB bandpowers are typically expressed
  with the normalization $\ell (\ell+1) C_{\ell}/(2\pi)$. For
  reference our result $C_{\ell}^{src} = 43.0 \nK^2$ corresponds to
  $X_{src} = 0.036 \jysq$, and at $\ell = 2500$, $\ell
  (\ell+1)C_{\ell}^{src}/(2 \pi) = 43.2 \uKbp$.} from the clean
simulations is $C_{\ell}^{src} = 44 \pm 14 \nK^2$; this is our best
estimate of the residual point source contamination in the CBI
fields. In comparison, the mean signal from all simulations
(neglecting the CBI/GBT orphan-source constraints) is $63^{+24}_{-30}
\nK^2$. The $95\%$ upper limits for the clean and total distributions
are $80 \nK^2$ and $204 \nK^2$, respectively.  The maximum power in
any of the 215 clean simulations is $112 \nK^2$, or almost exactly
half of what is needed to explain the excess power observed by CBI
over intrinsic anisotropy, compared $519 \nK^2$ in the total set of
simulation.  For the total set of simulations $2.2\%$ of instances
give power equal to or exceding what is needed to account for the CBI
high-$\ell$ excess. The long tail to high power in the total
distribution comes from {\it one to a few individual bright 30 GHz
sources} which would have been detected in the CBI maps were they
present.  The non-gaussian nature of the distribution is substantial:
the scatter in gaussian simulations with the same average power as the
clean simulations is a factor of $5.5$ lower.

%
%


Our observed level is in good agreement with, though generally lower
than, past measurements.  \citet{cbideep} found $0.08 \pm 0.04 \jysq
\, (C_{\ell}^{src} = 96 \pm 48 \nK^2)$, the value used in previous CBI
analyses, based on Owens Valley 40-m measurements.  The data in
\citet{cleary05} predict a mean level of $0.03 \, \jysq \,
(C_{\ell}^{src} = 36 \nK^2)$, with no uncertainty stated. These
results are consistent with what we report here.

\citet{sza} recently reported a determination of residual source power
$\ell (\ell +1) C_{\ell}^{src}/(2\pi) = 378 \pm 87 \uKbp$ following a
procedure similar to ours with data from the Sunyaev-Zel'dovich Array
(SZA). These results are also at $30$ GHz but on smaller angular
scales ($\ell \sim 4500$) where the discrete source power will be
higher.  Under the approximation that the flat-bandpower CMB window
functions adequately represent the effect of sources on the power
spectrum we can compare these results to the results of our
simulations above.  Our best estimate of the residual (sub-$3.4 \mJy$
in NVSS) source contamination at 31 GHz for {\it random} fields on the
sky is the total distribution neglecting the CBI map constraints,
resulting in a predicted $\ell (\ell +1) C_{\ell}^{src}/2\pi =
242^{+77}_{-97} \uKbp$ for the SZA measurement (including the extra
contribution estimated in \S~\ref{subsec:otherpops}).  Taking the SZA
error bar at face value these measurements are consistent at
$1.2\sigma$. The SZA error bar is based on Gaussian statistics so will
be a significant underestimate of the true uncertainty in the residual
source power spectrum.  The underestimate is a factor of $5.5$ for the
CBI fields but will be different for the SZA since the size of the
areas covered are very different. Conversely, were the entire CBI
excess to be explained by discrete sources the SZA should see $\ell
(\ell +1) C_{\ell}^{src}/(2\pi) = 875 \pm 221 \uKbp$. This is only
marginally ($2.1 \sigma$) consistent with the lower power level seen
by the SZA.


\subsubsection{The Contribution of sub-millijansky Galaxies}
\label{subsec:otherpops}

\begin{figure}[htbp]
\epsfxsize=6in
\epsffile{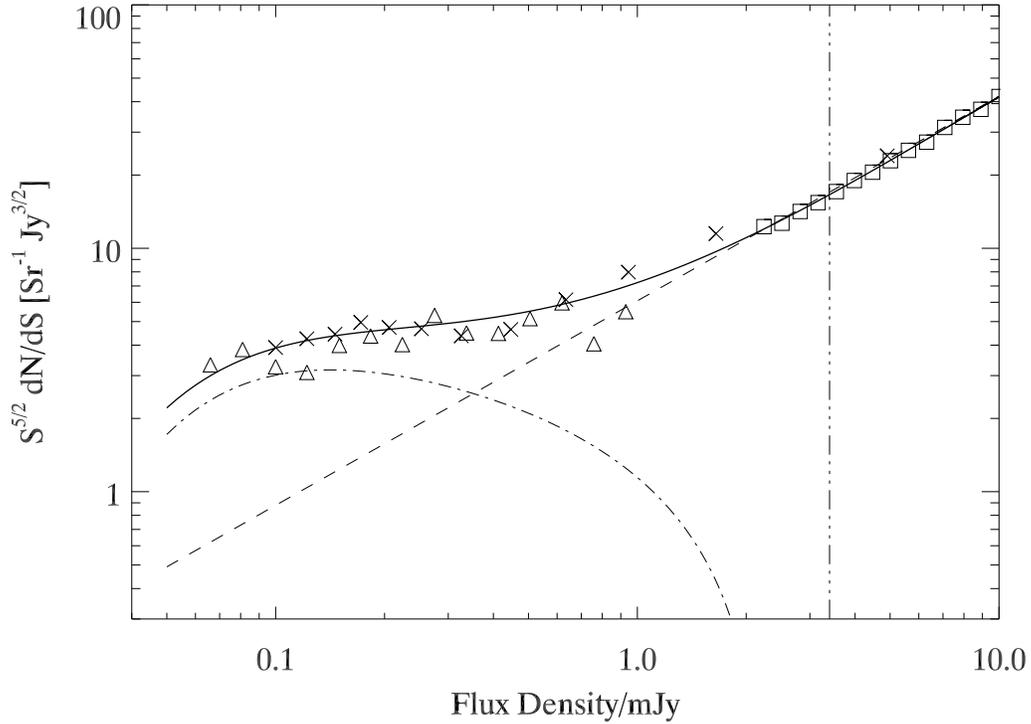}
\caption{$1.4$ GHz source counts. The solid line
is the model of \citet{hopkins03} and triangles are their
measurements in the Phoenix Deep Field; x's are the source
counts from the COSMOS field \citep{bondiCosmos08}; and
squares are source counts from FIRST \citep{white97}. The dashed
line is a power-law fit to the FIRST source counts at flux densities
fainter than $100 \, {\rm mJy}$ and the dot-dash line is the excess of 
the full source counts over the power law form. The vertical 
dash-triple-dot line is the CBI projection threshold: the sources of
interest are those to the left of this line. The power law behavior
persists up to $100$ mJy.}
\label{fig:dnds}
\end{figure}

Our simulations considered only the power-law-distributed source
population which is seen at mJy levels and higher in low frequency
surveys. We must also consider the contribution of fainter sources
likely belonging to a different population and having different
spectral properties.  At $1.4$ GHz flux densities under $\sim 1 \,
{\rm mJy}$ the source counts turn up due to the emergence of the
high-$z$ starbursting galaxies \citep[e.g.][]{windhorst85} --- see
Figure~\ref{fig:dnds}.  We can estimate the impact of this population
by first considering source correction at low frequency.  Explicitly
integrating the source count of \cite{hopkins03} from $50 \, {\rm \mu
  Jy}$ to $3.4 \, {\rm mJy}$ we find a total $1.4 \, {\rm GHz}$
residual source contribution of $C_{\ell}^{src} = 1038 \nK^2$. We
assume that the turn-up is due to a distinct population. Integrating
the power law over this range gives $869 \nK^2$, thus, the
contribution of the sources responsible for the turn-up in the counts
below a millijansky is $170 \nK^2$ at $1.4 \, {\rm GHz}$. This implies
that if every single one of these sources had a flat spectrum between
$1.4$ and $31$ GHz they would account for less than two-thirds of the
small-scale power in excess of the CMB observed by CBI of $270
\nK^2$. In reality, observations of $\mu {\rm Jy}$ sources
\citep{richards00} show typical spectral indices between $1.4$ and $8$
GHz of $-0.8$, consistent with the observed dominance of synchrotron
in nearby starbursting galaxies \citep{yuncarilli,condonreview}.

%

Assuming the distribution of spectral indices which we determined for
mJy level radio galaxies we obtain a 31 GHz value $C_{\ell}^{src} = 12
\nK^2$, a small correction to the mJy-AGN contribution of $44
\nK^2$. We include this contribution in the power spectrum analysis
for a final correction of $56 \nK^2$. \citet{dezotti05} also find that
sub-mJy galaxies make a minor contribution in comparison to mJy level
AGN.

It is possible that the sources responsible for the turn-up in the low
frequency counts could have a high incidence of inverted-spectrum
sources extending to 31 GHz thus contributing more to the high-$\ell$
source correction. Using simulations similar to those in
\S~\ref{subsec:srccounts} with modified spectral index distributions
we estimate that were these sources to have moderately inverted
spectral ($\alpha \sim 0.2$) they would need to constitute 40\% of the
sub-mJy population in order to fully explain the CBI excess. Were they
to have strongly inverted spectra ($\alpha \sim 0.8$), 2\% of the
population is required. In contrast, 
most steeply inverted spectrum source in the GBT+OVRO surveys the
had $\alpha = 0.49$ and $<0.1\%$
of sources had $\alpha > 0.3$.  Both cases would give rise to
substantial enhancements (factors of $1.5$ and $3$ for the strongly
and moderately inverted cases, respectively) enhancements of the 31
GHz source counts over those reported in \S~\ref{subsec:srccounts} in
the 1 to 10 mJy range.  


Preliminary analysis of deeper 31 GHz GBT \citep{masoninprep} and ATCA
\citep{taylorinprep} observations targetting a small sample of $\sim
40$ sources with $S_{1.4} \sim 1 \, {\rm mJy}$ indicates that the
mean $1.4-31$ GHz spectral index of these sources is comparable to
that of the AGN population and that there is not a substantial
population with inverted spectra continuing to 31 GHz.

\subsubsection{Source Clustering}
\label{subsec:clustering}

In addition to the Poisson or shot noise contribution of discrete
sources to the power spectrum there will also be a contribution due to
their spatial correlations, given by the second term of
\citep{scott99,oh03}
\begin{eqnarray}
C_{\ell}^{src} & \propto &
 \int_{0}^{s_{max}} \, ds \, s^2 \frac{dN}{ds} + 
          \omega_{\ell} \, \left(\int_{0}^{s_{max}} \, ds \, s \frac{dN}{ds}\right)^2 \\
 & \propto & \langle s^2\rangle  + \langle s \rangle^2 \, \omega_{\ell}
\end{eqnarray}
where $\omega_{\ell}$ are the coefficients of the Legendre polynomial
expansion of the discrete source angular correlation function (ACF)
$\omega(\theta)$.  The first term on the right hand side of this
equation is simply the shot noise contribution which has already been
considered in detail.  Using the \citet{blakewall} measurement of the
NVSS ACF, and the value $\langle S_{31} \rangle^2/\langle S_{31}^2
\rangle = 0.04$ determined from the 31 GHz flux density PDFs
determined in \S~\ref{sec:maxlike} we find that clustering is a
negligible contribution in comparison to the poisson term at these
faint flux densities.  The sub-mJy sources, which have a substantial
contribution from starbursting galaxies, could be more strongly
clustered than mJy AGN. Applying the $3\sigma$ upper limit of
\citet{webb03} on the clustering amplitude of submillimeter galaxies
to the sources lying above the AGN differential counts power law does
not change the conclusion from the calculation.


\subsubsection{Source Variability}
\label{subsec:variability}

Some sources exhibit significant time variability, with variability
measures increasing to timescales of up to $\sim 2 $ years
\citep{hughesAllerAller92}.  The effect of this will be to broaden our
measurement of the apparent $1.4$ to $31$ GHz flux density ratio
$r(t_1-t_2)=S_{31}(t_2)/S_{1.4}(t_1)$ over what would be observed in a
commensal multi-frequency survey, $r_0 = S_{31}(t_1)/S_{1.4}(t_1)$.
Provided the 31 GHz measurements are separated from the $1.4$ GHz
measurements by a period of time greater than the longest
characteristic timescale for variations, our measured distribution of
$r$ will be a statistically fair description of
$S_{31}(t_3)/S_{1.4}(t_1)$ for other 31 GHz occuring at some time
$t_3$ also separated from $t_1$ by greater than the longest
characteristic timescale for variation. Considering that the $1.4$ GHz
NVSS observations were collected from September 1993 October 1996, the
CBI observations from November 1999 to April 2005, and the GBT
observations from February to May of 2006, this will largely be the
case. There could be a small number of variable sources with
variability measures increasing beyond time spans of 10 years but
compared to the sample as a whole these are rare and will have little
impact on our results.

We note that while $r$ is a fair sample to describe the CBI residual
source power, extrapolations between frequencies other than $1.4$ and
$30$ GHz will be biased.  Even assuming that all sources
instantaneously have true power-law spectra, sources whose apparent
spectral indices fluctuate flatwards due to variability make very
different marginal contributions to the calculated sky variance at
another frequency than those whose apparent spectral indices fluctuate
steepwards. 

\section{Conclusions}
\label{sec:summary}

By measuring the 31 GHz flux densities of a large sample of $1.4$ GHz
selected sources we have for the first time characterized the 31 GHz
properties of a large sample of mJy-level radio galaxies.  Our sample
was large enough to place significant limits upon the frequency of
rare inverted spectrum sources which can contribute significantly to
the 31 GHz counts, and even more to the 31 GHz sky variance.  We find
the mean $31$ to $1.4$ GHz flux density ratio is $0.111 \pm 0.003$,
corresponding to a spectral index $-0.71 \pm 0.01$, and the mean
spectral index is $<\alpha>= -0.92^{+0.29}_{-0.30}$.  The fraction of
sources with $\alpha > -0.5$ is $9.0 \pm 0.8\%$ and the fraction with
inverted spectral indices $\alpha > 0$ is $1.2 \pm 0.2\%$.  This has
allowed us to greatly improve the accuracy with which we calculate the
statical point source correction for the Cosmic Background Imager
experiment.  We find that residual mJy-level AGN contribute a power of
$C_{\ell}^{src} = 44 \pm 14 \nK^2$. Including an additional estimated
$12 \nK^2$ contribution from faint ``sub-mJy'' sources, residual
sources account for $21 \pm 7 \%$ of the amplitude of the power seen
in excess of intrinsic anisotropy by CBI at $\ell > 2000$. We place a
95\% upper limit on residual source contamination of $C_{\ell}^{src} =
92 \nK^2$ or $34\%$ of the total excess power.  By way of comparison a
total residual point source correction $270 \pm 60 \nK^2$ is needed to
fully account for the observed CBI excess power. All of these results
are consistent (at $1.2\sigma$) with the recent SZA result of
\citet{sza}; a detailed comparison is given at the end of
\S~\ref{subsec:cbisims}.  Note that we express our results in terms of
$C_{\ell}$, which is appropriate for unresolved sources, rather than
$\ell (\ell+1) C_{\ell}/2\pi$.

A population of faint inverted-spectrum sources not present at
milli-Jansky levels could compromise these conclusions, but the
requirements are substantial: 20\% of $S_{1.4} < 1 \mJy$ sources would
need to have $\alpha_{1.4-30} = +0.2$ for instance, a factor of twenty
more than is observed at mJy levels in this survey; or 2\% of $S_{1.4}
< 1 \mJy$ sources would need to have $\alpha_{1.4-30} = +0.8$,
resulting in $\sim 8$ of these sources per square degree above a
milli-Jansky at 30 GHz.  In this survey of $3,165$ sources only a
single source was as steeply inverted as $\alpha = 0.5$. Both
scenarios would imply enhancements of the 31 GHz source counts over
what we have measured by at least 50\% at 1 mJy.


It is worth noting several points in connection with these
conclusions.  First, it is essential to appreciate that the residual
sources are fundamentally selected by $1.4$ GHz flux density. Second,
for an unbiased calculation of the 31 GHz contribution of these
sources a complete $1.4$ to $31$ GHz (effective) spectral index
distribution must be used. Populations {\it selected} at a higher
frequency will have preferentially flatter spectral indicies; and
spectral indices measured between $1.4$ and a higher frequency less
than $31$ GHz will not in general be representative, and in
particular, will not reflect the steepending of synchrotron spectral
indices to higher frequencies. Both of these effects, due to the large
lever arm in frequency involved, have a significant impact. Third, it
is essential to avoid selection biases in estimating the spectral
index distribution. In the absence of better information previous CBI
results used an incomplete sample of $1.4$ to $31$ GHz spectral
indices (biased flat) resulting in an overestimate of the point source
contribution.  The Bayesian analysis of \S~\ref{sec:maxlike}
eliminates any bias due to censoring (non-detections) at 31 GHz.
Fourth, as discussed in \S~\ref{subsec:variability}, spectral index
extrapolations from other frequencies are biased by source
variability.  Finally it is important to use an accurate form of the
well-known low-frequency counts rather than simple approximations to
them.  All of these considerations are independent of the 31 GHz
counts {\it per se}, which are only indirectly related to the
conclusions reached: to calculate the CBI residual source correction from
31 GHz counts requires the same additional information (the
distribution of $S_{31}/S_{1.4}$) as calculating the statistical
correction from the $1.4$ GHz counts.

We have also computed 31 GHz counts based on $1.4$ GHz source counts
and our distribution of spectral indices, finding $N(>S_{31}) = 16.7
\pm 1.7 \times (S_{31}/{\rm 10 mJy})^{-0.80 \pm 0.07} \, {\rm
  deg^{-2}}$ for $1 \mJy < S_{31} < 4 \mJy$, in good agreement with
observed 31 GHz source counts at higher flux densities, as well as the
model of \citet{dezotti05}.

The National Radio Astronomy Observatory is a facility of the National
Science Foundation operated under cooperative agreement by Associated
Universities, Inc. We thank the GBT and OVRO science and engineering
staff for outstanding contributions to both survey projects and
acknowledge support from NSF grants AST-9413935, AST-9802989,
AST-0098734, and AST0206416. We thanks Gianfraco deZotti for providing
us with his most recent 30 GHz source count model; Dan Marrone for
providing the SZA window functions; Jim Condon, Bill Cotton, Mike
Jones, and Angela Taylor for helpful discussions; and Rachel Rosen for
carefully proofreading the manuscript.  We thank Liz Waldram and Guy
Pooley for providing unpublished source size information from the 9C
survey. Finally we thank an anonymous referee for thorough comments
which helped to improve the paper.


\bibliography{cbi11}

\appendix

\section{Confusion Correction}
\label{appendix:confusion}

Consider that we have a set of $N_{obs}$ potentially mutually confused
measurements $S_{obs,i}$ and seek to determine the true flux densities
$S_{T,i}$ of the targetted sources in the presence of a number of
other contributing $N_{est}$ sources for which we have no
measurements, but only some uncertain flux density estimates
$S_{est,j}$. These quantities are related by
\begin{eqnarray}
\left( \begin{array}{c}
             S_{obs,1} \\
             S_{obs,2} \\
              ...      \\
             S_{obs,Nobs}
       \end{array}         \right) & = &
                                      \left(\begin{array}{cccc}
                                            B_{11} & B_{12} & ... & B_{1,Nobs} \\    
                                            B_{21} & B_{22} & ... & B_{2,Nobs} \\ 
                                                   & ...    &     &  \\
                                            B_{Nobs,1} & ... &    & B_{Nobs,Nobs} \\ \end{array} \right)
  \left( \begin{array}{c}
             S_{True,1} \\
             S_{True,2} \\
              ...      \\
             S_{True,Nobs}
       \end{array}         \right)        +  ... \\ \nonumber
 & & \left(\begin{array}{cccc}					      
      B_{11} & B_{12} & ... & B_{1,Nest} \\    		      
      B_{21} & B_{22} & ... & B_{2,Nest} \\ 		      
             & ...    &     &  \\				      
      B_{Nobs,1} & ... &    & B_{Nobs,Nest} \\ \end{array} \right)
  \left( \begin{array}{c}
             S_{est,1} \\
             S_{est,2} \\
              ...      \\
             S_{est,Nest}
       \end{array}         \right)                                      
\label{eq:confusedsrcs}
\end{eqnarray}
where the $B_{ij}$ is the beam weight that source $j$ contributes to
observation $i$, for a Gaussian beam 
\begin{equation}
B_{ij} = exp\left( -
\frac{|\vec{x_i}-\vec{x_j}|^2}{2 \sigma_{beam,i}^2}\right)
\end{equation}
for an observation pointed at $\vec{x_i}$ and a source at
$\vec{x_j}$. $\sigma_{beam,i}$ is the Gaussian beam width for
measurement $i$, potentially different for different measurements if
the data come from different telescopes as is the case here. This is
a straightforward system of $N_{obs}$ equations in $N_{obs}$ unknowns.

We must also account for our off-source (or reference) beam
positions. Due to our position-nodded beam-switching observing
strategy, and the fact that the beamswitching occurs in azimuth, the
reference (off-source) beams sweep out arcs on the sky giving a
greater chance of encountering a second source; but tracking over a
range of parallactic angles will cancel this effect on average.  The
OVRO data, with observations of a given source spread over many days
or even months, sample a wide range of parallactic
angles. Consequently for the OVRO data we use the average reference
beam arcs in calculating the $B_{ij}$.  The GBT data, in contrast,
typically have a single measurement at a well-defined parallactic
angle for a given source.

Using the reference catalog we scan the full set of OVRO observations
and identify those for which the sum of the absolute values of
beam-weighted confusing source flux densities in the reference catalog
amounts to more than half the measurement error for that source.  We
solve Eqs.~\ref{eq:confusedsrcs} for this subset of observations and
use the corrected values in the catalog; revised uncertainties are
also calculated by propagating the measurement errors, assuming
$\sigma(S_{est}) = S_{est}$ for the non-measured sources (determined
from the distribution of $S_{31}/S_{1.4}$ of \S~\ref{sec:maxlike}).
For the OVRO data this affects $1.3\%$ of the data. 

\section{Maximum Likelihood}
\label{appendix:maxlike}

We can rewrite the likelihood in terms of flux
density ratios and us the fact that the measurement errors are uncorrelated to get:

\begin{equation}
P\left(\frac{\skaobs}{\slobs} \right) \propto \int d\skatrue P\left(\skaobs
| \skatrue\right) \int P\left(\frac{\skatrue}{\sltrue}|\sltrue\right)
P\left(\frac{\sltrue}{\slobs}\right)  d\sltrue
\end{equation}
The outer integral is a convolution with the 31 GHz (OVRO and GBT)
measurement error distribution; the inner integral folds the flux
density ratio through the $1.4 $ GHz measurement error distribution,
including a Malmquist-bias correction in the $P(\slobs | \sltrue)$
which assumes differential source counts of the form
$S^{-1.66}$. Since the measurements of each source are independent,
the total log-likelihood is the sum of the log-likelihoods for
individual sources, and evaluating the likelihood reduces to
(efficiently) evaluating the individual PDFs for all the 31 GHz flux
density measurements (OVRO and GBT). We assume that $\sratio$ is
independent of the $1.4$ GHz flux density over the range
of interest, approximately 1 to $3.4$ mJy.

Several points should be noted:
\begin{enumerate}
\item Under the assumption (to be checked) that the $\sratio$ distribution
is independent of the $1.4$ GHz flux density over the range of interest,
we can take $ P\left(\frac{\skatrue}{\sltrue}|\sltrue\right)
\rightarrow  P\left(\frac{\skatrue}{\sltrue}\right)$ \ie\ 
the distribution is the same for all sources and needs to be
calculated only once.
\item For computational efficiency, we pre-calculate the FFT of the
  inner integral, $P\left({\frac{S_{31,T}}{S_{1.4,obs}}}\right)$
for a range of $1.4 $ GHz signal-to-noise values at each likelihood step, then linearly
interpolate between these values for a given source.
\item Calculating $P(\skaobs | \slobs)$ requires convolving $P(S_{31,true})$ with the 31 GHz
noise.  We use the fact that we wish to evaluate
$P(S_{31,observed})$ only at the actual observed flux density to speed up the
convolution.  Because we calculate the Fourier transform of
$P(S_{31,true}|S_{1.4,observed})$ in the previous step, the
convolution with the (Gaussian) 31 GHz noise is an anlytic
multiplication.  We evaluate the back-transform only at the two
FFT points bracketing the observed flux density, then linearly interpolate
between them.  This saves us from having to do a full FFT for each
source, with computational load reduced from $N\log(N)$ to $2N$.

\item Because differences in log-likelihood are meaningful, the
  maximum-likelihood process naturally produces meaningful errors in
  the flux density distribution that
  fold in both the measurement errors and the uncertainty due to the
  finite number of sources measured.

\end{enumerate}
We parameterize $\sratio$ with a set of evenly spaced points in the
frequency spectral index $\alpha$, and do a piecewise-Hermite cubic
polynomial interpolation through them, padded with zeros on either
side.  The Hermite interpolation is similar to a spline, but has two
key advantages: the interpolation is always local so it never rings,
and the function values between interpolation points are always
bracketed by the function values at points.  So, as long as our model
points stay non-negative, the interpolated PDF will also be strictly
non-negative.

\end{document}